\documentclass[]{aa}  

\usepackage{graphicx}
\usepackage[version=4]{mhchem}
\usepackage{txfonts}
\usepackage{float}
\usepackage{subcaption}
\usepackage[colorlinks=true,urlcolor=blue,citecolor=blue,linkcolor=blue,breaklinks=true]{hyperref}
\newcommand{\subsubsubsection}[1]{\paragraph{#1}\mbox{}\\}
\usepackage{xcolor}

\defcitealias{Oberg2022}{Paper I}

\begin{document} 

   \title{Circumplanetary disk ices}

   \subtitle{II. Composition}

   \author{N. Oberg
          \inst{1,2}
          \and
          S. Cazaux
          \inst{2,3}
          \and
          I. Kamp 
          \inst{1} 
          \and
          T.-M. Bründl
          \inst{2,3}
          \and
          W.F.Thi
          \inst{4}
          \and
          C. Immerzeel
          \inst{2}
          }

   \institute{Kapteyn Astronomical Institute, University of Groningen, P.O. Box 800, 9700 AV Groningen, The Netherlands \\
              \email{oberg@astro.rug.nl}
         \and
             Faculty of Aerospace Engineering, Delft University of Technology, Delft, The Netherlands 
         \and
             University of Leiden, P.O. Box 9513, 2300 RA, Leiden, The Netherlands     
         \and 
             Max-Planck-Institut für extraterrestrische Physik, Giessenbachstrasse 1, 85748 Garching, Germany \\
             }

   \date{Received --- accepted ---}

 
  \abstract
   {The subsurface oceans of icy satellites are among the most compelling  among the potentially habitable environments in our Solar System.  The question of whether a liquid subsurface layer can be maintained over geological timescales depends on its chemical composition. The composition of icy satellites is linked to that of the circumplanetary disk (CPD) in which they form. The CPD accretes material from the surrounding circumstellar disk in the vicinity of the planet, however, the degree of chemical inheritance is unclear.}
   {We aim to investigate the composition of ices in chemically reset or inherited circumplanetary disks to inform interior modeling and the interpretation of in situ measurements of icy solar system satellites, with an emphasis on the Galilean moon system.}
   {We used the radiation-thermochemical code ProDiMo to produce circumplanetary disk models and then extract the ice composition from time-dependent chemistry, incorporating gas-phase and grain-surface reactions.}
   {The initial sublimation of ices during accretion may result in a CO$_2$-rich ice composition due to efficient OH formation at high gas densities.  In the case of a Jovian CPD, the sublimation of accreted ices results in a CO$_2$ iceline between the present-day orbits of Ganymede and Callisto. Sublimated ammonia ice is destroyed by background radiation while drifting towards the CPD midplane. Liberated nitrogen becomes locked in N$_2$ due to efficient self-shielding, leaving ices depleted of ammonia. A significant ammonia ice component remains only when ices are inherited from the circumstellar disk.}
   {The observed composition of the Galilean moons is consistent with the sublimation of ices during accretion onto the CPD.  In this scenario, the Galilean moon ices are nitrogen-poor and CO$_2$ on Callisto is endogenous and primordial.  The ice composition is significantly altered after an initial reset of accreted circumstellar ice.  The chemical history of the Galilean moons stands in contrast to the Saturnian system, where the composition of the moons corresponds more closely with the directly inherited circumstellar disk material.}


   \maketitle
%

\section{Introduction}

The search for habitable worlds beyond the solar system has historically focused on planets in the so-called "habitable zone," where surface conditions theoretically support the presence of liquid water \citep{Hart1979}.  In the Solar System, however, icy satellites and minor bodies outside of the classical habitable zone are the most common type of worlds that are known to host oceans of liquid water \citep{Hussmann2006,Nimmo2016}. Evidence strongly supports the presence of a subsurface ocean on the Galilean satellites Europa and Ganymede, as well as (to a lesser extent) Callisto  \citep{Carr1998,Khurana1998,Kivelson2002,Sohl2002,Saur2015}. The resonant configuration of the satellites prevents a damping of the orbital eccentricities, producing levels of tidal heating capable of sustaining subsurface oceans over geological timescales \citep{Peale593,Hussman2004,SHOWMAN1997367}. Whether or not a given level of tidal heating produces subsurface melt depends in part on the composition of the satellite ices.  The proposed abundant impurities include NH$_3$, CH$_4$, CO, and CO$_2$, along with salts MgSO$_4$ and NaCl \citep{KARGEL1992,Mousis2006}.  The liquidus temperature of co-deposited ice mixtures can be depressed by the presence of NH$_3$  \citep{Choukroun2010,Sohl2010} or methanol (CH$_3$OH) \citep{Deschamps2010,Dougherty2018}, as well as salts to a lesser extent.  Hence, the composition of the volatile reservoir from which icy satellites form is of direct relevance to the presence of a subsurface ocean, their geothermal and physical  evolution \citep{Hammond2018}, the interpretation of in-situ geophysical measurements \citep{Vance2018}, and the eventual atmospheric composition by outgassing or impact dissociation \citep{Sekine2014,Glein2015}.

%


In particular, ammonia is important to the interior state and evolution of icy bodies. The presence of NH$_3$ in the form of dihydrate can drive differentiation of rock and ice  \citep{Desch2009}. Ammonia in a pure H$_2$O-NH$_3$ eutectic system produces a freezing point depression of $\sim$100$\,$K \citep{KARGEL1992,Grasset2000,Leliwa2002}.  Ammonia can also reduce the density of melt with implications for buoyancy and cryovolcanism  \citep{Croft1988}, while increasing viscosity and reducing the efficiency of convection \citep{Grasset2000}.   Ammonia has been detected in the plumes of Enceladus \citep{Waite2009} but not on the surface of the Galilean moons.  Tentative evidence for a subsurface ocean on Callisto would be bolstered by the presence of an ammonia component of 1-5$\%$ \citep{kirk1987,Showman1999, Spohn2003}. 


In the "gas-starved" circumplanetary disk (CPD) paradigm, moon formation occurs in a relatively low-mass, cool disk that must accumulate solids to form giant moons over time \citep{Canup2002,2006Natur.441..834C,Batygin2020}.  Infalling material from the surrounding circumstellar disk may be shock-heated in the process of accretion onto the CPD \citep{Szulagyi2017,Szulagyi2017b,Aoyama2018}, with increasing shock temperature for increasing planetary mass.  If the shock heating chemically resets infalling gas or ices, new ice formation must occur within the CPD to produce the icy satellites we see today.  The resulting composition of the satellite ices may then depart substantially from those in the planetary feeding zone.

Prior works modeling equilibrium condensation chemistry in a Jovian CPD suggest that in the event of an initial vaporization of ices, the ``mostly inefficient" gas-phase reactions lead to ratios of CO$_2$:CO:CH$_4$ and N$_2$:NH$_3$ that are not substantially different from those in the feeding zone of Jupiter \citep{Mousis2006,Mousis2006b}.  However, it has long been recognized that grain-surface chemistry plays a critical role in the formation of many common molecules under interstellar conditions \citep{Hasegawa1992,Dishoeck1998,Cazaux2002, Caselli2004, Garrod2006b, Ruaud2015,WAKELAM2017B}. The use of a more comprehensive modeling approach including grain-surface and photochemistry to revisit the formation of ices in CPDs is thus motivated.  We aim to investigate the composition of ices that form in a chemically reset CPD with viscous timescale $10^3-10^4$ yr, where infalling ices are sublimated and gas is atomized by shock-heating. These results will be contrasted with a partial reset in which only ices are sublimated during accretion and with a full chemical inheritance scenario in which the composition of the circumstellar disk gas and ice is preserved.  We intend to link observations of solar system icy satellites with modern chemical disk models to lay the foundation for our understanding of how icy moons are built up from material in the CPD.

\section{Methods}

We used the radiation-thermochemical disk modeling code \textsc{ProDiMo}\footnote{https://prodimo.iwf.oeaw.ac.at/} to model gas and dust chemistry and physics in disks    \citep{Woitke2009,Woitke2016, Kamp2010,Kamp2017,Thi2011,Thi2020H2}. The gas-grain chemistry is solved self-consistently with the 2D radiative transfer and heating and cooling balance using a rate equation-based approach.  Most reaction rates are selected from the UMIST2012 database \citep{McElroy2013} and  three-body collider reactions are adopted from the UMIST2006 rate file \citep{Woodall2007}, as they were not included in the 2012 release. In the following sections we review the implementation of the grain surface chemistry (Sect. \ref{sec:methods:surfchem}), extensions to our standard chemical network (Sect. \ref{sec:methods:nondiana}), and properties of the CPD model (Sect. \ref{sec:methods:cpdmodel}).  

\subsection{Grain surface chemistry} \label{sec:methods:surfchem}

\textsc{ProDiMo} includes a rate-equation based, statistical two-phase approach to gas and dust grain surface chemistry that is largely based on the work of \citet{Hasegawa1992}. Gas-phase atoms and molecules can become weakly adsorbed to grain surface physisorption sites. Physisorbed species diffuse in a random-walk process "hopping" from one physisorption site to another \citep{Barlow1976}. Diffusion occurs thermally if there is sufficient energy to overcome a diffusion barrier or can otherwise occur by tunneling. The surface diffusion rate is the sum of the thermal, tunneling, and cosmic-ray induced diffusion rates. The rate of thermal diffusion of species, $i,$ is:

\begin{equation}
    R_i^{\rm diff,th} = \nu_{0,i} Q_i^{\rm diff}(a_i^{\rm diff},E_i^{\rm diff}) e^{\Delta E_{ij}/k_{\rm B}T_{\rm d}} / n b_{\rm site} \, \textrm{s}^{-1}
,\end{equation}
\noindent
where $k_{\rm B}$ is the Boltzmann constant, $T_{\rm d}$ is the dust temperature, and the frequency term, $\nu_{\rm 0, i}$, describing characteristic lattice vibration is:

\begin{equation}
v_{0, i}=\sqrt{\frac{2 N_{\mathrm{surf}} E_{i}^{\mathrm{b}}}{\pi^{2} m_{i}}},
\end{equation}
\noindent
 $N_{\rm surf}$ is the surface density of adsorption sites \mbox{1.5$\times10^{15}$ cm$^{-2}$}.  The number of adsorption sites per monolayer on a grain of radius $a$ is \mbox{$nb_{\rm site} = 4\pi N_{\rm surf} a^2$}. Also, $E^{\rm b}_{\rm i}$ is the binding energy and $m_i$ is the mass of the species.  We use a tunneling-corrected form of the Arrhenius equation, Bell's formula, $Q_{i}^{\rm diff}$, to determine the surface diffusion tunneling rate \citep{Bell1980}:

\begin{equation}
    Q_{i}^{\rm diff} = \frac{\beta e^{-\alpha} - \alpha e^{-\beta}}{\beta - \alpha},
\end{equation}
\noindent 
where
\begin{equation}
    \alpha = E_i^{\rm diff} / k_{\rm B} T_{\rm d}
\end{equation}
\noindent
and
\begin{equation}
    \beta = \frac{4 \pi a_i^{\rm diff}}{h} \sqrt{2 m_i E_i^{\rm diff}}.
\end{equation}
\noindent 
Here, $E_i^{\rm diff}$ is the diffusion activation energy, $a_i^{\rm diff}$ is the diffusion activation barrier width, and $h$ is the Planck constant. $\Delta E_{ij}$ is the difference in binding energy between two adsorption sites, $i$ and $j$. We have $\Delta$ E$_{ij}$ = 0 if \, E$_i^b \leq$ E$_j^b$, thus, for a random hop between two physisorption sites, $\Delta E = 0$.  The cosmic-ray induced diffusion rate $R_i^{\rm diff,CR}$ is adopted from \citep{Hasegawa1993,Reboussin2014}:


\begin{equation}
    R_i^{\rm diff,CR} = f(70 \textrm{K}) R_i^{\rm diff,th}(70 \textrm{K}) \frac{\zeta_{\rm CR}}{5 \times 10^{-17}} \textrm{s}^{-1}
,\end{equation}
\noindent
where $\zeta_{\rm CR}$ is the cosmic ray ionization rate and $f(70 \rm K) = 3.16\times10^{-19}$ is the duty-cycle of the grain at 70 K. The total diffusion rate, $R_i^{\rm diff}$, is then the sum of the thermal and cosmic-ray induced diffusion rates.

Species that are physisorbed to a grain surface can react directly with species coming from the gas-phase (Eley-Rideal) or with other physisorbed species (Langmuir–Hinshelwood). Reaction products can remain on the grain surface or chemically desorb into the gas-phase. Reactions between physisorbed species follow the prescription of \citet{Hasegawa1992}.  The reaction rate coefficients between two surface-adsorbed species is the probability of a reaction per encounter multiplied by the encounter rate between the two species diffusing across the surface.  The encounter rate between two adsorbed species $i$ and $j$ hopping across the surface is then:

\begin{equation}
    k_{ij} = \kappa_{ij}(R_i^{\rm diff}+R_j^{\rm diff})/n_d \,\,\, \textrm{cm}^3 \textrm{s}^{-1}
,\end{equation}
\noindent
where $\kappa_{ij}$ is the reaction probability, $R_i^{\rm diff}$ and $R_j^{\rm diff}$ are the diffusion rates (s$^{-1}$) for species, $i$ and $j$, and $n_{\rm d}$ is the dust grain number density (cm$^{-3}$). The reaction probability, $\kappa_{ij}$, takes into account the competition between association of the species and diffusion \citep{Garrod2011,Bonfanti2016,Ruaud2016}:

\begin{equation}
    \kappa_{ij} = \frac{Q_{\rm Bell}(a_{ij}^{r},E_i^{\rm act})}{Q_{\rm Bell}(a_{ij}^{r},E_i^{\rm act}) + P_i^{\rm diff} + P_j^{\rm diff}}
,\end{equation}
\noindent
where $a^{r}_{ij}$ is the reactive barrier width,  $E_i^{\rm act}$ is the activation energy of the reaction barrier, and $P_i^{\rm diff}=R_i^{\rm diff}/ \nu_{0,i}$.  


We assume the semi-equilibrium theory in which reactions between physisorbed and gas-phase species (Eley-Rideal) is equal to the probability of the gas atom colliding with the physisorbed species multiplied by the probability of the gas-phase species having sufficient energy to overcome the reaction barrier. Impinging gas-phase species are assumed to have an energy relative to the surface species $1/2 k_{\rm B}T_{\rm g} + E_i^b$, where $T_{\rm g}$ is the gas temperature and $E_i^b$ is the binding energy. Photon and cosmic-ray induced dissociation and desorption of grain-surface species are also included. Adsorption and desorption processes are described fully in \citet{Thi2020H2}. 


\subsubsection{Extending chemistry beyond the standard network} \label{sec:methods:nondiana}

We developed an extended chemical network based on the \mbox{"DIANA standard large"} network described in \citet{Kamp2017}, which contains 235 species (including 63 ices) and 12 elements + polycyclic aromatic hydrocarbons (PAHs) and is optimized for gas-phase chemistry + direct adsorption and desorption from grains.  The use of grain-surface reactions necessitates the inclusion of several additional species to the DIANA standard chemical network to capture the relevant chemistry occurring at the disk midplane. These seven additional gas-phase species and six additional ices are listed in Table \ref{tab:extra-species}. Physisorbed atomic hydrogen and H$_2$ are included for their critical role in many grain-surface reactions.  Hydrogenated PAH and O$_2$H are included for their relevance to the chemical reset scenario in which atomic H is initially very abundant. The rationale for their inclusion is discussed in the following sections. In addition, HOCO and HCOOH are more directly involved in the formation of relevant ices and their roles are discussed in Sect.  \ref{sec:results_CO2}.

\begin{table}
    \caption{Non-standard species included in the chemical network.}

    \centering
    \renewcommand{\arraystretch}{1.1}%

   \begin{tabular}{lll}
    \hline \hline 
     gas-phase species & \\ \hline
      O$_2$H             &         \\
      HOCO  & \\
      HCOOH & \\ 
      HCOOH$^+$ & \\ 
      HCOOH$_2^+$      &         \\
      PAH-H & \\
      PAH-H$^+$  &      \\ \hline
      ices      & E$_{\rm ads}$ [K]  \\ \hline
      H\#       &  600$^1$           \\
      H$_2$\#   &  430$^2$           \\
      O$_2$H\#  &  3650$^2$          \\
      HOCO\#    &  2000$^3$          \\
      HCOOH\#   &  5000$^4$          \\
      PAH-H\#   &  5600$^5$          \\

    \end{tabular}
     \vspace{1ex}

     {\raggedright Adsorption energies are adopted from $^1$\citet{Cazaux2002}, $^2$\citet{Garrod2006}, $^3$\citet{Ruaud2015}, $^4$\citet{Oberg2009} $^5$\citet{Thrower2009} \par}

    \label{tab:extra-species}
    
\end{table}

\subsubsubsection{Hydrogenated polyclic aromatic hydrocarbons (PAH-H)}

In \textsc{ProDiMo,} the formation rate of H$_2$ can be calculated in multiple ways.  The standard approach is that H$_2$ formation proceeds via a pseudo-reaction at a rate calculated according to the analytical approach of \citet{Cazaux2002}m which presupposes that surface-chemisorbed H atoms play a dominant role at high temperatures ($\geq$100\,K).  However, the formation of H$_2$ is calculated explicitly when grain-surface reactions are included in the reaction network \citep{Thi2020H2}.  It was noted in the accompanying work that H$_2$ formation occurs in parallel with H$_2$O ice deposition on grains at the midplane when the CPD is chemically reset (\citet{Oberg2022} hereafter, Paper I).  The formation of H$_2$O ice after the reset is rapid and a median-sized grain is coated in several ($\gg$3) monolayers of water ice prior to the complete conversion of H to H$_2$. This poses a problem as the formation of H$_2$ via surface-chemisorbed H is considered implausible when the number of water ice monolayers exceeds a certain number ($\sim3$) \citep{WAKELAM2017B}.  We assume that the diffusion timescale of the atomic hydrogen in a water ice matrix and subsequent difficulty of H$_2$ escape from the grain silicate surface precludes this formation pathway in our scenario. An alternative path to form H$_2$ is via hydrogenated polycyclic aromatic hydrocarbons (\mbox{PAH-H}) \citep{Bauschlicher1998}.  Experimental and theoretical works have demonstrated that H$_2$ can form via Eley-Rideal abstractions on neutral PAHs \citep{Bauschlicher1998, Rauls2008, Mennella2012, Thrower2012} and cationic PAHs \citep{Hirama2004, Cazaux2016, Boschman2012}.    We include in the chemical network the singly hydrogenated species PAH-H, PAH-H$^+$, and the physisorbed ice form PAH-H$\#$ \citep{Thrower2009} to enable this formation path.

As a first step towards H$_2$ formation, the neutral or ionized PAH is hydrogenated with a small (324 K) activation barrier. The H$_2$ formation at the CPD midplane then proceeds primarily via 

\begin{align}
    & \rm \operatorname{PAH-H} + H \rightarrow H_2 + PAH, 
\end{align}
\noindent
and to a lesser extent ($\sim 1-10\%$ of the total H$_2$ formation rate depending on location in the CPD), directly via the gas-phase neutral-neutral reactions:

\begin{align}
& \rm  H + HCO \rightarrow H_2 + CO, \\
& \rm  H + HNO \rightarrow H_2 + NO.
\end{align}

While we do include several grain-surface reactions to form H$_2$ (e.g., H$\#$ + HCO$\#$ $\rightarrow$ CO$\#$ + H$_2\#$, O$\#$ + H$_2$CO$\#$ $\rightarrow$ CO$_2\#$ + H$_2\#$); in practice, these occur at a negligible rate due to the 50 K minimum dust temperature in the CPD. The resulting efficiency of H$_2$ formation is lower than the analytic rate of \citet{Cazaux2002} in part due to the low ambient temperatures (< 200 K, which in combination with the activation barrier impede the process) at the optically thick midplane. The correspondingly longer time over which atomic hydrogen is present has direct consequences for the efficiency of water formation. Gas-phase H$_2$O can then form via the hydrogenation of OH for an extended period of time (discussed further in \citetalias{Oberg2022}).

\subsubsubsection{O$_2$H}

The hydroperoxyl radical O$_2$H is a very reactive oxygen species that we have found to play a role in the formation of methanol in the inner region of chemically reset CPDs.   We include the gas and ice form of O$_2$H in the extended chemical network, with an adsorption energy of 3650 K \citep{Garrod2006}.  The oxygen-bearing gas-phase species abundances are sensitive to the presence of O$_2$H at high densities.  Three-body collider reactions with free atomic hydrogen and O$_2$ produce O$_2$H. This reaction has been extensively studied both theoretically \citep{Horowitz1985, Sellevag2008, Morii2009} and experimentally \citep{Kurylo1972, Davidson1996,Hahn2004,Mertens2009} at high and low temperatures.  With the inclusion of O$_2$H in the extended network the gas-phase O$_2$ reservoir at the midplane  (nominally present at an abundance  $\sim$10$^{-4.4}$ relative to hydrogen in the standard network) is depleted and converted via OH into H$_2$O through the following reactions 

\begin{align}
  &  \rm O_2 + H + M \rightarrow  O_2H + M,     
\end{align}
\noindent
or 
\begin{align}
  &  \rm O_2 + HCO \rightarrow  O_2H + CO,     
\end{align}
\noindent
followed by
\begin{align}
  &  \rm O_2H + H    \rightarrow  OH + OH,      \\
  &  \rm OH   + H    \rightarrow  H_2O + photon,      
\end{align}
\noindent
These reactions compete for the free H that is required to form methanol via 
\begin{align}
  &  \rm H_2CO + H \rightarrow \rm CH_3O  ,\\
  &  \rm CH_3O + H \rightarrow CH_3OH,          
\end{align}
\noindent
and thus suppress its formation. The inclusion of O$_2$H in the chemical network reduces the abundance of methanol ice interior to the NH$_3$ iceline relative to the results of the standard chemical network by 90-99$\%$. However, this has a negligible impact on  the total disk-integrated methanol abundance.

\subsection{Circumplanetary disk model} \label{sec:methods:cpdmodel}

We adopted the properties of the \textsc{ProDiMo} circumplanetary disk model developed in \citetalias{Oberg2022}. The CPD is a "gas-starved," actively fed accretion disk \citep{Canup2002} that is heated primarily by viscous dissipation at the midplane \citep{,dAlessio1998,Frank2002}.   The parameters of the reference CPD model are listed in Table \ref{tab:cpds}. The physical, radiative, and thermal properties of the CPDs are demonstrated in \citetalias{Oberg2022}, namely, Figures 3 and 4.

The disk structure in terms of radial and vertical dimension, density profile, dust-to-gas ratio, and temperature, are assumed to exist in a steady-state and are kept fixed. Following the gas-starved disk paradigm, the CPD does not  instantaneously contain the solid mass required to form the Galilean satellites. The total refractory dust mass is 1.7$\times10^{-5}$ M$_{\oplus}$ and exists in the form of small grains (0.05-3000 \textmu m). The dust grain size distribution is described by a smooth power-law $n \propto a^{-3.5}$.  Such a disk is optically thick out to approximately $\sim$1/3rd of the planetary Hill radius $R_{\rm H}$, which is coincident with the theoretical limit for stable orbits  \citep{1998ApJ...508..707Q, 10.1111/j.1365-2966.2009.15002.x, 2011MNRAS.413.1447M}. 

For the ice-to-rock ratio of the solids in the CPD to be consistent with the ice-to-rock ratio of the outer Galilean satellites, it was found in \citetalias{Oberg2022} that the dust-to-gas ratio of the CPD may be depleted relative to the canonical $10^{-2}$ by a factor $\gtrsim10-20$.  This depletion in dust corresponds with the rapid inwards drift and loss of grains larger than $\sim$150 \textmu m, which was found to occur naturally for a disk with a mass of $10^{-7}$ M$_{\odot}$ and accretion rate $\dot M = 10^{-11}$M$_{\odot}$yr$^{-1}$ \citepalias{Oberg2022}. Alternatively, pressure-bump trapping at the gap edge can directly prevent larger grains from accreting onto the CPD \citep{Rice2006,Morbidelli2012,Zhu2012,Bitsch2018}. For assumptions regarding the efficiency of settling, surface density power-law, and maximum grain size,  it was found in \citetalias{Oberg2022} that the global dust-to-gas ratio should not exceed $10^{-3.3}$ for a CPD with a mass of \mbox{$10^{-7}$ M$_{\odot}$} and accretion rate of \mbox{$\dot M$ = $10^{-11}$ M$_{\odot}$yr$^{-1}$} to satisfy the constraints on Jovian icy moon bulk composition. The properties of the disk models are further justified and detailed in \citetalias{Oberg2022}, where the authors explored a small grid of plausible CPD parameters. 

In this work, our analysis is focused on the case of the 10$^{-7}$\,M$_{\odot}$ CPD with accretion rate $\dot M = 10^{-11}$M$_{\odot}$yr$^{-1}$, as it was most closely able to reproduce the radial ice-to-rock ratio of the Galilean satellites while being consistent with the circumstellar disk gas component expected lifetime.  We contrast our results with those of the ``high" viscosity, hotter CPD with $\dot M = 10^{-10}$M$_{\odot}$yr$^{-1}$ and correspondingly shorter viscous timescale $10^3$ yr given uncertainties in the magnitude of the disk viscosity. 

\subsubsection{Initial chemical conditions}

Three different initial chemical conditions are considered. The first case is a full chemical ``reset,"  in which ices are initially sublimated and the circumstellar gas is reverted to a purely atomic, ionized state. A chemical reset may occur if, for instance, the circumstellar material is shock-heated during accretion onto the CPD, if the gas and dust are irradiated while crossing the optically thin gap, or if material only flows into the gap from the upper optically thin surface layers of the circumstellar disk.  The CPD model is initialized in this fully reset state after which it is allowed to chemically evolve over its viscous timescale $t_{\rm visc}$. The viscous timescale of the disk is defined as the time over which the majority of the gas mass is lost  $t_{\rm visc}$ = M$_{\rm cpd} / \dot M,$ where $\dot M$ is the mass accretion rate.  We assume that gas is lost to viscous radial flow either to decretion beyond the disk edge or accretion onto the planet.  As the disk mass is assumed to be constant, the net inflow-outflow rate of matter is necessarily zero.  Our reference CPD model has a viscous timescale of $10^4$ yr with a corresponding midplane heating rate equivalent to an $\alpha$-viscosity of $10^{-3.6}$.  We contrast these results with a ``partial reset" in which only the ices are placed back in the gas-phase.  This is similar to the work of \citet{Mousis2006} wherein the authors consider a case in which infalling ices are initially sublimated in a warm disk which subsequently cools, although we consider a disk with a static temperature structure.  Finally we consider an ``inheritance" case in which the chemical composition at the circumstellar disk outer edge is used as the initial state. The abundance of the most common species for these three initial conditions can be found in Appendix \ref{appendix:initial}. The circumstellar disk model and the sampling of the inheritance chemistry are described in the accompanying  \citetalias{Oberg2022}. 

\begin{figure}
\centering
  \includegraphics[width=0.45\textwidth]{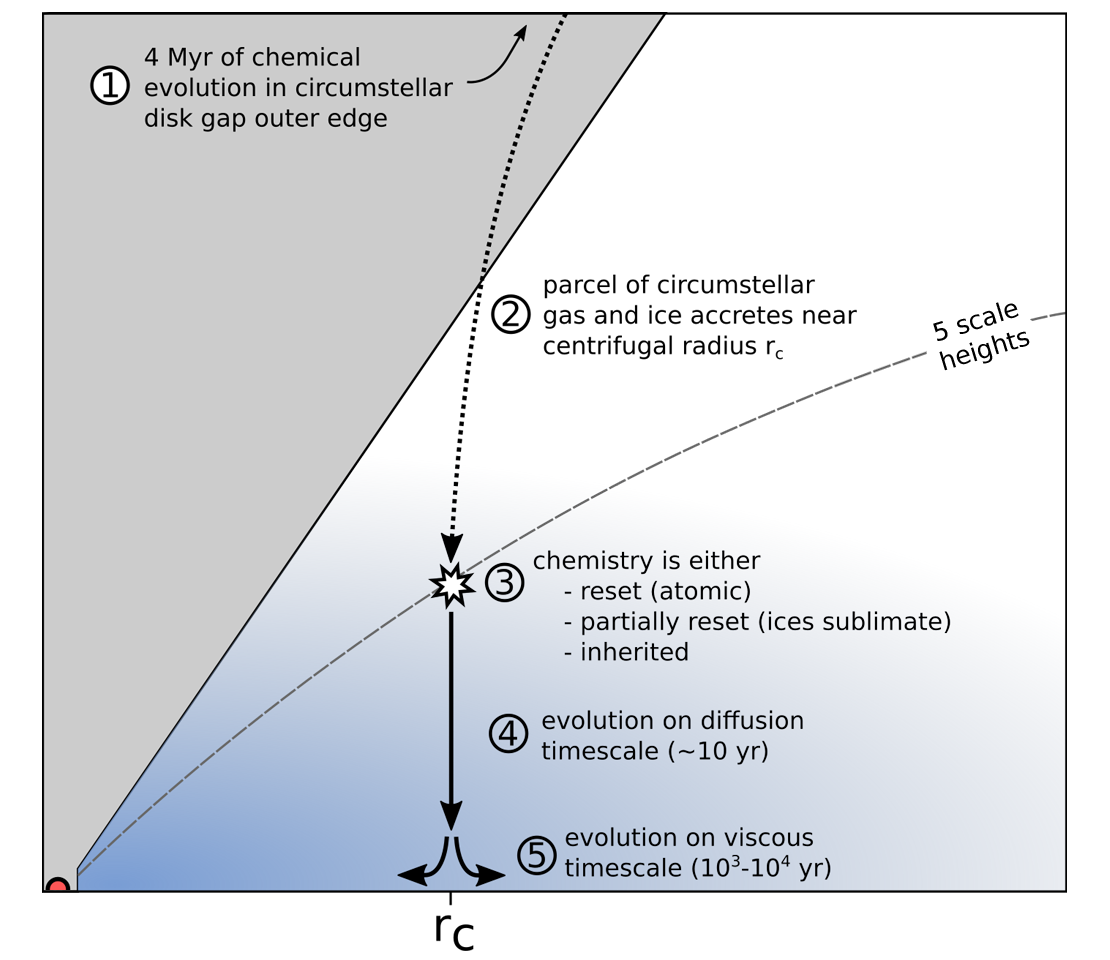}
      \caption{Schematic illustration of the modeling process.  In \mbox{step 1,} the chemistry in a circumstellar disk model is evolved for \mbox{4 Myr}. This chemistry is extracted from the gap outer wall region and used as a starting point prior to accretion.  To consider various possible accretion scenarios, the composition of the infalling material is either reset to atomic (full reset), the ices are sublimated (partial reset) or the chemistry remains unaltered (inherit). In \mbox{step 2,} the chemistry of a parcel of gas and ice is evolved for 10 yr as it travels towards the CPD midplane.  In step 3, the chemistry is evolved at the CPD midplane for the viscous timescale of the disk.}
      \label{fig:chemdiagram}
\end{figure}

It is also necessary to consider the consequences of the gas and dust being shocked at several scale heights above the CPD midplane \citep{Takasao2021} prior to the gas turbulently diffusing downwards into the optically thick region.  The ambient conditions at $\sim$5 pressure scale heights ($A_{\rm V}$ = 0.01) differ significantly from those at the midplane  ($A_{\rm V}$=21) given the magnitude of the external stellar irradiation.  To take into account this gradual change in ambient conditions, we incorporated an additional step necessary to prevent the sublimated ices immediately re-adsorbing to grains.  We adapted the model to follow a single parcel of gas and dust that is initialized above the midplane and then settles towards the midplane at the centrifugal radius ($\sim0.03$ R$_{\rm H}$) \citep{2008ApJ...685.1220M}.  This process is labeled as step 2 in \mbox{Fig.\ref{fig:chemdiagram}}.   

In this step, we evolved the chemistry in a 0D grid-cell for a fraction of the diffusion timescale.  The resulting composition of the gas and ice was extracted and used to populate a new grid-cell, in which the background conditions are updated to correspond to the downwards motion of the gas parcel. The extracted relative species abundances were simply applied to the new cell and absolute abundances were rescaled to correspond to the new grid-cell density.  This process was repeated iteratively as ambient conditions (optical depth, density, and gas and dust temperature) change.  As a simplification owing to significant uncertainties in the origin, magnitude, and spatial distribution of turbulence within the CPD, we simply assumed that the parcel travels at a constant rate until it reaches the midplane. The timescale of this process is $\sim 10$ yr \citepalias{Oberg2022}, although this value is still highly uncertain. Accordingly, we also considered diffusion timescales of 1, 10, and 100 yr. The final composition of the parcel at the midplane was then used to populate the CPD midplane for the final step (step 3 in Fig.\ref{fig:chemdiagram}), whereby chemical evolution proceeds up until the viscous timescale.

\begin{table}
    \caption{Parameters of the reference CPD model. }

    \centering
    \renewcommand{\arraystretch}{1.1}%

   \begin{tabular}{lll}
    \hline \hline
        Parameter               & Symbol          & Value                 \\ \hline
        Planetary mass             & $M_{\rm p}$     & 1.0 M$_{\rm J}    $   \\
        Planetary luminosity    & $L_{\rm p}$     & $10^{-5}$ L$_{\odot}$   \\
        
        Effective temperature   & $T_{\rm eff,p}$      & 1000 K           \\
        UV luminosity           & $L_{\rm UV,p}$       & 0.01 L$_{\rm p}$$^{*}$  \\
        Interstellar UV field   & $\chi$               &  $3\times10^3$   \\ 
        Background temperature  & $T_{\rm back}$       & 50 K             \\

        \hline

        Disk mass                 & $M_{\rm cpd}$      & 10$^{-7}$ M$_{\odot} $       \\
    
        Disk inner radius         & $R_{\rm in,cpd} $  & 0.0015 au                  \\
        Exponential decay radius  & $R_{\rm in,cpd} $  & 0.11 au                    \\ 
        Disk outer radius         & $R_{\rm out,cpd}$  & 0.34 au                    \\
        Column density power ind. & $\epsilon$         & 1.0                        
\\ \hline
        Accretion rate            &  $\dot M$   & $10^{-11}$-$10^{-10}$ M$_{\odot}$ yr$^{-1}$  \\
        Viscosity                 &  $\alpha$   & $10^{-3.6}$-$10^{-2.7}$                     \\
        
        \hline
        Minimum dust size      & $a_{\rm min}$    & 0.05 \textmu m       \\
        Maximum dust size      & $a_{\rm max}$    & 3000 \textmu m        \\
        Dust-to-gas ratio      & $d/g$            & $10^{-3.3}$  \\       
        Flaring index          & $\beta$          & 1.15               \\
        Reference scale height & $H_{\rm 0.1 au}$ & 0.01 au            \\

    \end{tabular}
\caption*{\textbf{*} Planetary UV luminosity is expressed in multiples of the planetary luminosity,  L$_{\rm p}$.  }
    \label{tab:cpds}
    
\end{table}

\subsection{Likelihood of chemical reset and magnitude of shock-heating} \label{sec:shock-heating}

\begin{figure}
  \includegraphics[width=\textwidth/2]{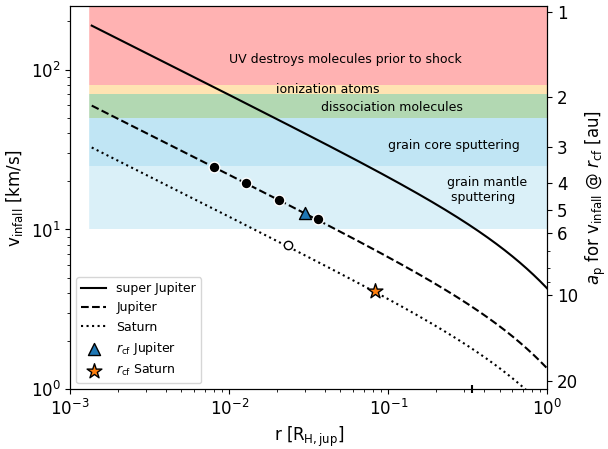}
      \caption{Velocity of material falling onto a CPD $v_{\rm infall}$ at radius $r$ for planets of Saturnian (dotted line), Jovian (dashed lines), and super-Jovian (10 M$_{\rm J}$) (solid line) mass.  The centrifugal radii $r_{\rm cf}$ of Jupiter and Saturn are indicated by the blue triangle and orange star, respectively. The radial position of the Galilean satellites is indicated by the four black circles and the radial position of Titan is indicated by the white circle. The planetary orbital radius, $a_{\rm p}$, that corresponds to the infall velocity at the centrifugal radius, $v_{\rm infall}(r_{\rm cf}),$ is indicated on the right vertical axis. The shaded colored regions indicate different chemical consequences of shock-heating corresponding to a given $v_{\rm infall}$. The units on the abscissa are in multiples of the Jovian Hill radius.  All calculations correspond to a solar mass star.}
      \label{fig:shocking}
\end{figure}

Icy grains passing through an optically thin gap at 5 au around a Sun-like star can retain their icy mantles if swept up by the planet within $\sim10-100$ orbital timescales \citepalias{Oberg2022}.  If a (partial) chemical reset occurs, it must instead be due to either accreted material originating from a higher altitude in the circumstellar disk where ices are unstable or, otherwise, shock-heating on the CPD surface. We can estimate the shock velocity of infalling matter where it strikes the CPD and consider which of our initial chemical conditions corresponds most appropriately to the formation of the Galilean moon system.  Angular momentum of infalling circumstellar gas and dust causes it to accrete onto the CPD  near the so-called centrifugal radius, $r_{\rm cf}$ \citep{Hayashi1985,2008ApJ...685.1220M}. The infall velocity at $r_{\rm cf}$ must be $\gtrsim 8-10$ km s$^{-1}$ for dust grain icy mantles to be lost due to sputtering and thermal desorption \citep{Woitke1993,Aota2015,tielens2021}.  We approximated the infall velocity as a function of planetocentric radius by considering orbits with apoapsis of a single circumstellar disk pressure scale height at the position of Jupiter ($z = 0.5$\,au) \citepalias{Oberg2022}, with orbital eccentricities corresponding to passage through the planet equatorial plane at some distance $r$. The resulting infall velocities, $v_{\rm infall}$, can be seen in Fig. \ref{fig:shocking} for planets of Saturnian, Jovian, and super-Jovian (10 M$_{\rm J}$) mass.  The infall velocity at $r_{\rm cf}$ is independent of the planetary mass, but it is instead a function of the planetary semimajor axis (for a circular orbit).  The shock velocity at the centrifugal radius of Jupiter is in the regime of icy mantle loss to sputtering \citep{Draine1993,Tielens2005}.  Hence, if the majority of grains accrete near $r_{\rm cf}$, Jupiter's CPD may be best represented by the ``partial reset" scenario. Conversely, no ice sublimation is expected to occur due to shock-heating in the case of Saturn. A full chemical reset is more likely to occur for a super-Jupiter at a stellocentric distance of 2-3 au from a solar-mass star.

\subsection{Chemical network diagrams}

Throughout this work, we make use of algorithmically generated chemical network diagrams to describe relations between atomic and molecular species, their relative abundances, formation rates, and the types of reactions that are involved.  The diagrams are generated with an implementation of the PyVis software package (itself based on the VisJS library \citet{Perrone2020}).  A description of how these diagrams are generated and interpreted can be found in Appendix \ref{appendix:diagrams}.

\section{Results and discussion} \label{sec:results}

\begin{figure}
  \includegraphics[width=\textwidth/2]{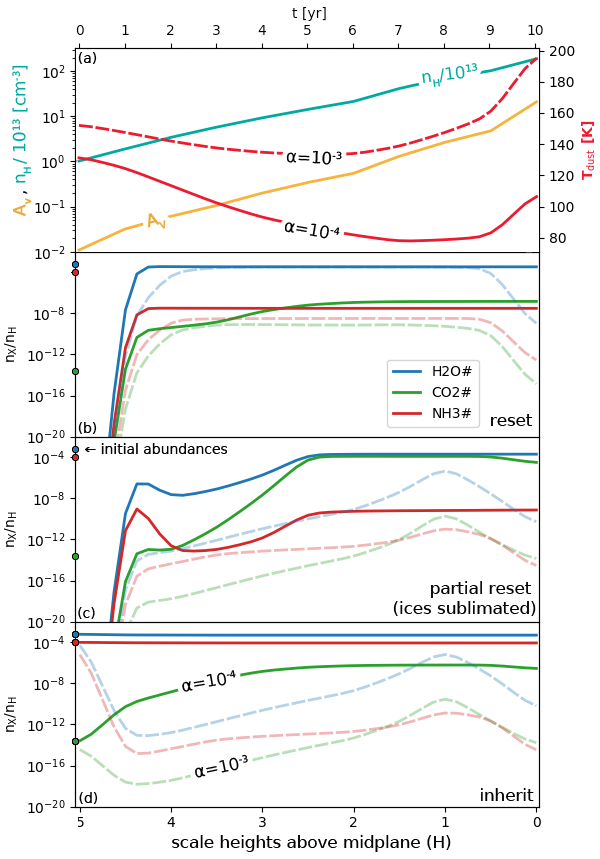}
      \caption{(a) Properties of a vertical slice in the circumplanetary disk  at the centrifugal radius, from five pressure scale heights (left) to the midplane (right).  Evolution of the abundance of H$_2$O, NH$_3$ and CO$_2$ ice as the parcel of gas and dust sinks to the midplane after accretion in the reset case: (b) in the partial reset case (c) and the inheritance case (d).  Solid lines trace the relevant properties and abundances for the low viscosity ($t_{\rm visc} = 10^4$ yr) case and the dashed line describes the high viscosity ($t_{\rm visc} = 10^3$ yr) case.}
      \label{fig:descent-evolution}
\end{figure}

Prior to reaching the midplane, the accreted gas and dust diffuses downwards from the optically thin surface layer of the CPD at the centrifugal radius, $r_{\rm c}$. We iteratively evolved the disk chemistry as the background conditions change during the descent.  The relevant properties of the vertical slice through the circumplanetary disk during the descent to the midplane at $r_{\rm c}$ can be found in Fig. \ref{fig:descent-evolution} panel (a). The post-shock evolution of the water ice abundance during the descent to the midplane can be found in panels (b), (c), and (d) of Fig. \ref{fig:descent-evolution} for the reset, partial reset, and inheritance cases, respectively.  Solid lines trace the evolution of ice impurity abundances as the gas parcel moves downwards from five scale heights (left) to the midplane (right).  Dashed lines trace the abundances in the case of a hotter, higher viscosity CPD with $t_{\rm visc}=10^3$ yr.  The initial pre-shock abundances of the impurities are indicated by the colored circles on the ordinate.

In the case of the reference (low-viscosity) fully or partially reset CPD, significant quantities of water ice have already formed prior to reaching the midplane.  In the fully reset case, the ice is predominantly water with $<0.1\%$ impurities in the form of CH$_3$OH and HCOOH ice.  In the partial reset case, a significant (25$\%$) CO$_2$ component has formed.  In the inheritance case, ices are able to survive the brief exposure to the optically thin upper layers of the disk and the CPD accretes a nitrogen-rich ice composition. In the high-viscosity ($\alpha=10^{-3}$) CPD model, ices are not thermally stable at the centrifugal radius midplane. This can be seen in \mbox{Fig. \ref{fig:descent-evolution},} where ice abundances decline immediately prior to reaching the midplane.   Consequently the initial post-shock conditions of a "partial reset" and "inheritance" converge to a similar ice-free molecular gas composition by the time the gas parcel reaches the midplane.

\begin{figure*}[!htp]
\centering

\begin{subfigure}{\textwidth/3}
    \centering\captionsetup{width=.8\linewidth}%
    \scalebox{0.85}{\textbf{Mass fraction of impurities}}\par\medskip \vspace{-1.5ex}
    \scalebox{0.85}{\textbf{and ice-to-rock ratio}}\par\medskip 
    \includegraphics[width=\textwidth]{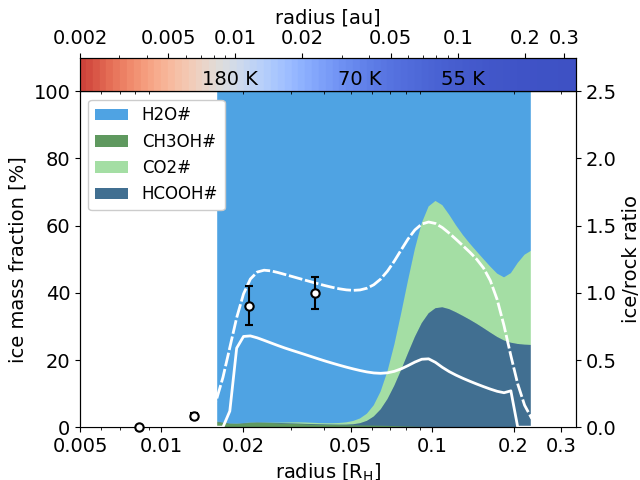}
             \caption{full reset}
\end{subfigure}\hfill%
\begin{subfigure}{0.25\textwidth}
    \centering\captionsetup{width=.8\linewidth}%
   \scalebox{0.85}{\textbf{Molar fraction of impurities}}\par\medskip
    \includegraphics[width=\textwidth]{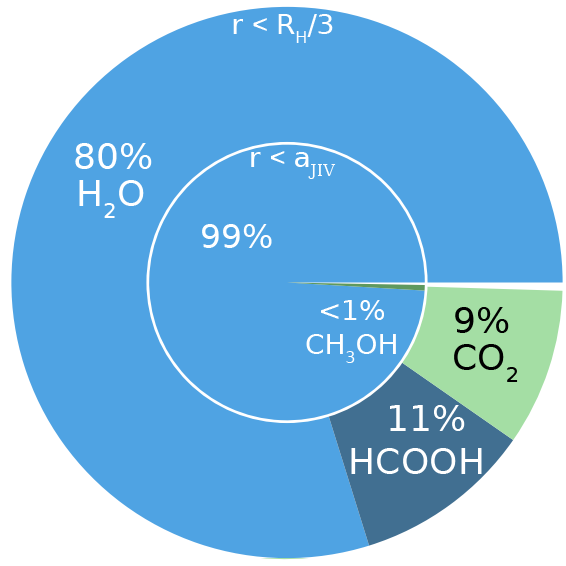}
             \caption{}
\end{subfigure}\hfill%
\begin{subfigure}{0.25\textwidth}
    \centering\captionsetup{width=.8\linewidth}%
    \scalebox{0.85}{\textbf{Elemental composition of the ice}}\par\medskip
    \includegraphics[width=\textwidth]{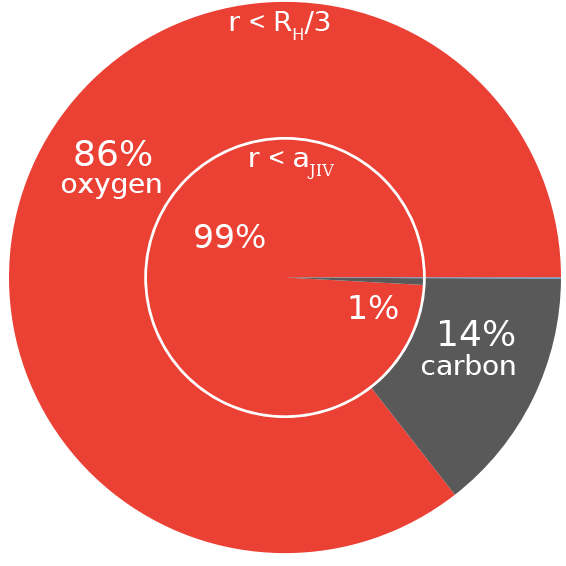}
             \caption{}
\end{subfigure}

\begin{subfigure}{\textwidth/3}
    \centering\captionsetup{width=.9\linewidth}%
    \includegraphics[width=\textwidth]{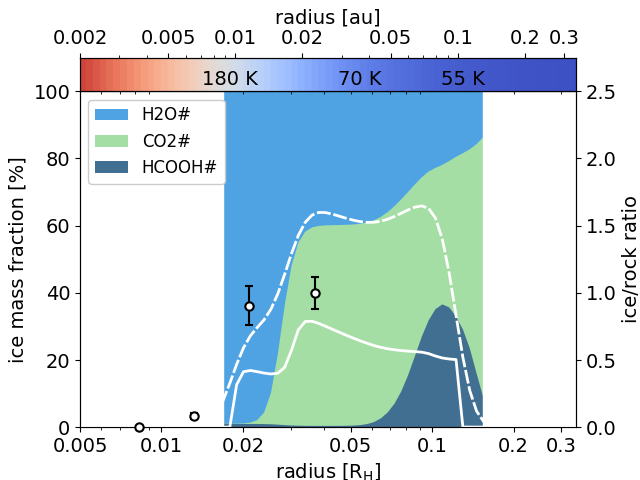}
             \caption{partial reset (only ices reset)}
\end{subfigure}\hfill%
\begin{subfigure}{0.25\textwidth}
    \centering\captionsetup{width=.9\linewidth}%
    \includegraphics[width=\textwidth]{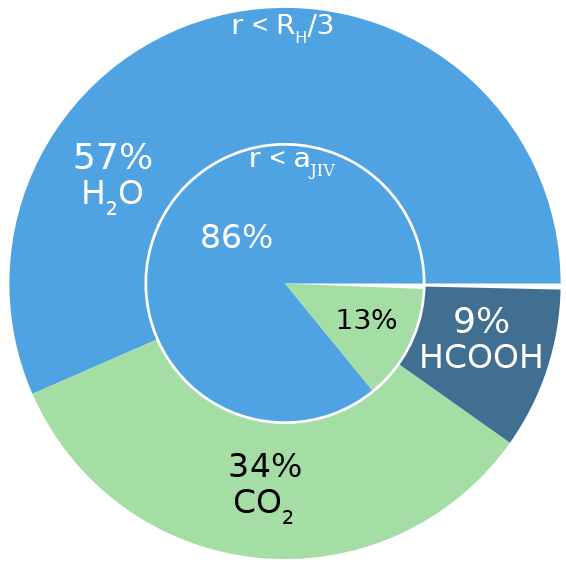}
             \caption{}
\end{subfigure}\hfill%
\begin{subfigure}{0.25\textwidth}
    \centering\captionsetup{width=.9\linewidth}%
    \includegraphics[width=\textwidth]{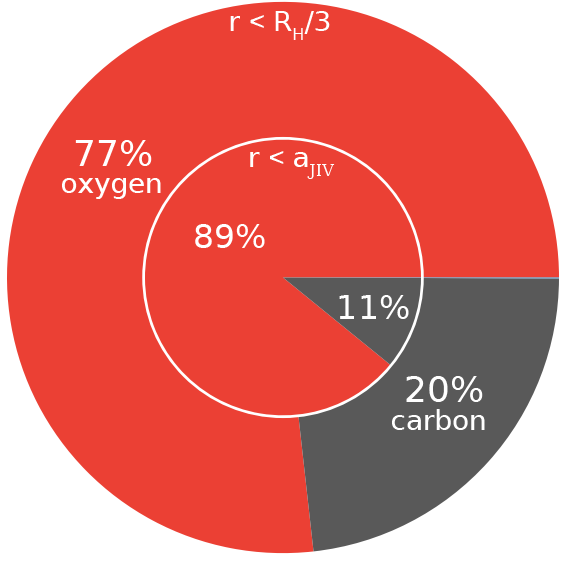}
             \caption{}
\end{subfigure}

\begin{subfigure}{\textwidth/3}
    \centering\captionsetup{width=.9\linewidth}%
    \includegraphics[width=\textwidth]{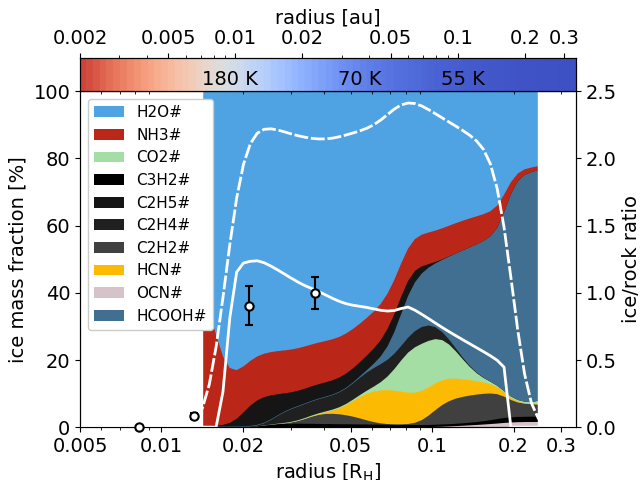}
             \caption{full inheritance}
\end{subfigure}\hfill%
\begin{subfigure}{0.25\textwidth}
    \centering\captionsetup{width=.9\linewidth}%
    \includegraphics[width=\textwidth]{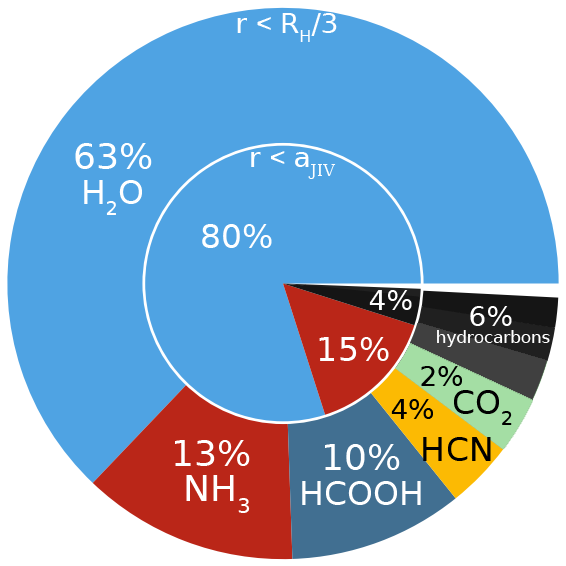}
             \caption{}
\end{subfigure}\hfill%
\begin{subfigure}{0.25\textwidth}
    \centering\captionsetup{width=.9\linewidth}%
    \includegraphics[width=\textwidth]{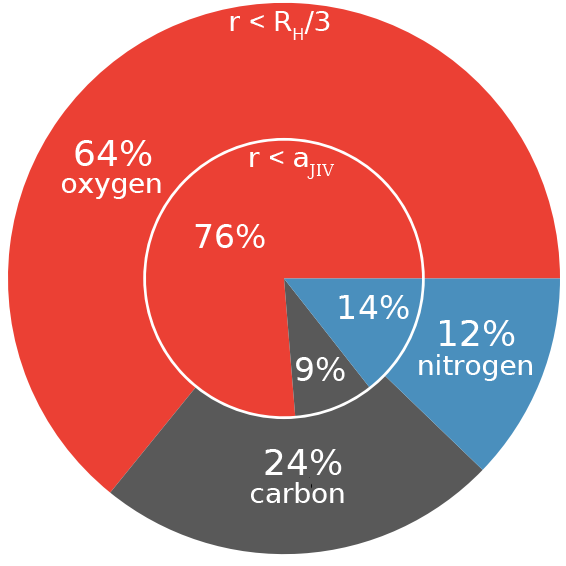}
             \caption{}
\end{subfigure}

\caption[short]{Overview of the chemical composition at the CPD midplane for the ``full reset" case (\textit{top row}), for the ``partial reset" case (\textit{middle row}), and for the ``full inheritance" case (\textit{bottom row}).  \textit{Left column:} Radial mass fraction of ices at the CPD midplane (filled-colored regions) where f$_{\rm ice}>0.01$. The white lines indicate the radial ice-to-rock ratio of solids at the midplane (solid line) and integrated up to an altitude above the midplane equal to the Hill radius of Ganymede (dashed line). The estimated ice-to-rock ratio of the Galilean satellites is included (circles with error bars). \textit{Center column}: Radially integrated midplane ice composition out to $R_{\rm H}/3$ (outer ring) and within the orbit of Callisto a$_{\rm JIV}$  (inner circle). \textit{Right column}: Total disk-integrated elemental composition of the ices are shown in the same two radial zones. }

\label{fig:detailed-composition-and-pie}

\end{figure*}

\begin{figure} 

  \includegraphics[width=\textwidth/2]{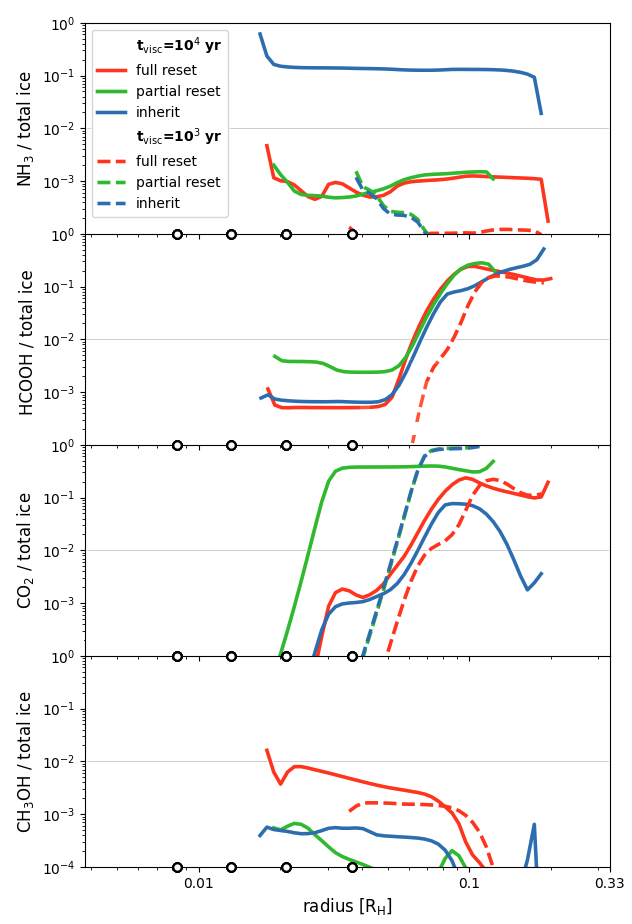}

\caption[short]{Radial abundance of selected non-H$_2$O ices as a fraction of the total ice abundance for the low-viscosity case with $t_{\rm visc} = 10^4$ yr (solid lines) and high-viscosity case with $t_{\rm visc} = 10^3$ yr (dashed lines). The position of the Galilean satellites are indicated by the empty circles. A light gray horizontal line indicates a concentration of 1$\%$.} \label{fig:radial-abundances-all}

\end{figure}

After the step involving the accreted gas and dust being followed as it travels towards the midplane (step 4 in \mbox{Fig. \ref{fig:chemdiagram}}), the resulting chemical abundances are used to specify the initial conditions for the rest of the CPD as it evolves on the viscous timescale (step 5 in \mbox{Fig. \ref{fig:chemdiagram}}). After 10$^3$-10$^4$ yr of further evolution, we extracted the radial ice composition at the midplane from six distinct CPD models describing three initial chemical conditions (reset, partial reset, and inheritance) and two disk $\alpha$-viscosities (corresponding to viscous timescales of 10$^3$ and $10^4$ yr).  An overview of the radial midplane ice composition,  the disk-integrated total molecular ice composition, and the disk-integrated total elemental ice budget of the low-viscosity CPDs can be found in Fig. \ref{fig:detailed-composition-and-pie}. For reference, the ice-to-rock ratio of the solids at the CPD midplane is also included in Fig.\ref{fig:detailed-composition-and-pie} (left column) as a solid white line.  The settling of large grains to the midplane strongly reduces the local ice-to-rock ratio.  Realistically, accreting moons may be able to capture solids drifting at higher altitudes above the midplane within their gravitational sphere of influence.  Hence, we included also the ice-to-rock ratio of solids integrated up to an altitude equal to the Hill radius of a Ganymede-mass object (dashed white line). The radial abundance profiles of NH$_3$, HCOOH, CO$_2$, and CH$_3$OH ices can be found in Fig. \ref{fig:radial-abundances-all}.

Ices at the partially or fully reset CPD midplane are found to contain significant impurities in the form of CO$_2$ and HCOOH, as well as, to a lesser extent, CH$_3$OH.  The chemically inherited CPD additionally contains HCN and hydrocarbon ices which were already present at the time of accretion.  Trace amounts of OCN, SO, SO$_2$, NH, NH$_2$, OH, and HNO ices can also be found, but each at $<0.1-0.5\%$ of the total ice mass. Although several of these ices have negligible absolute abundances, the fraction of their key element which has frozen out can be substantial. In particular, sulfur has frozen completely out of the gas-phase outside of the centrifugal radius in all cases.  The element fraction in ice can be found in Appendix \ref{appendix:freezeout}. In the following subsections, we discuss the formation and abundance of the impurities NH$_3$, CO$_2$, HCOOH, and CH$_3$OH.

\subsection{Partial reset (initial sublimation of ices)} \label{sec:results:partial_reset}
      
It is likely that in the case of the Jupiter system, the shock velocity of matter accreting at the centrifugal radius did not lead to the full dissociation of molecules.  A less extreme C-type shock-heating could simply cause icy grain mantles to desorb by sputtering, for instance. Accordingly, we focus our analysis and discussion on this case, whereby all ices are put back into their respective gas-phase counterpart.

\subsubsection{Ammonia (NH$_3$)} \label{sec:results_ammonia}

Immediately after accretion onto the CPD and the sublimation of ices, hydrogen is predominantly found in the form of H$_2$, oxygen in H$_2$O, and nitrogen in NH$_3$ and HCN at a ratio 1:0.63.  After ten years of drifting towards the midplane the gas is still H$_2$-dominated, but nitrogen is found primarily in N$_2$.   After being initially sublimated a minor fraction of the NH$_3$ immediately re-adsorbs to the grains (see Fig. \ref{fig:descent-evolution} (c)), but it is not stable against photodissociation given the background UV field intensity ($\chi_{\rm RT} > 1000$).  Above 2-3 scale heights, the NH$_3$ ice is photodissociated on the grain surface to, for instance, NH$_2\#$ and H$\#$ or back into the gas phase as NH.  Once the majority of nitrogen is locked into N$_2$ via NH+NH, it is stable against photodissociation due to self-shielding \citep{Li2013}, preventing the accumulation of NH$_3$. The photodissociation timescale of N$_2$ is much larger than the disk viscous timescale.   

Near the midplane, NH$_3$ ice forms by direct adsorption from the gas phase onto dust grains. The gas-phase NH$_3$ originates primarily via a sequence of three body collider reactions:

\begin{align}  \label{eq:nh3-collider-path}
    & \rm H_2 + N + M \rightarrow NH_2 + M ,\\ 
    & \rm NH_2 + H + M \rightarrow NH_3 + M.  \label{eq:NH2_collider_2}
\end{align}

\begin{figure}
\centering
  \includegraphics[width=\textwidth/2]{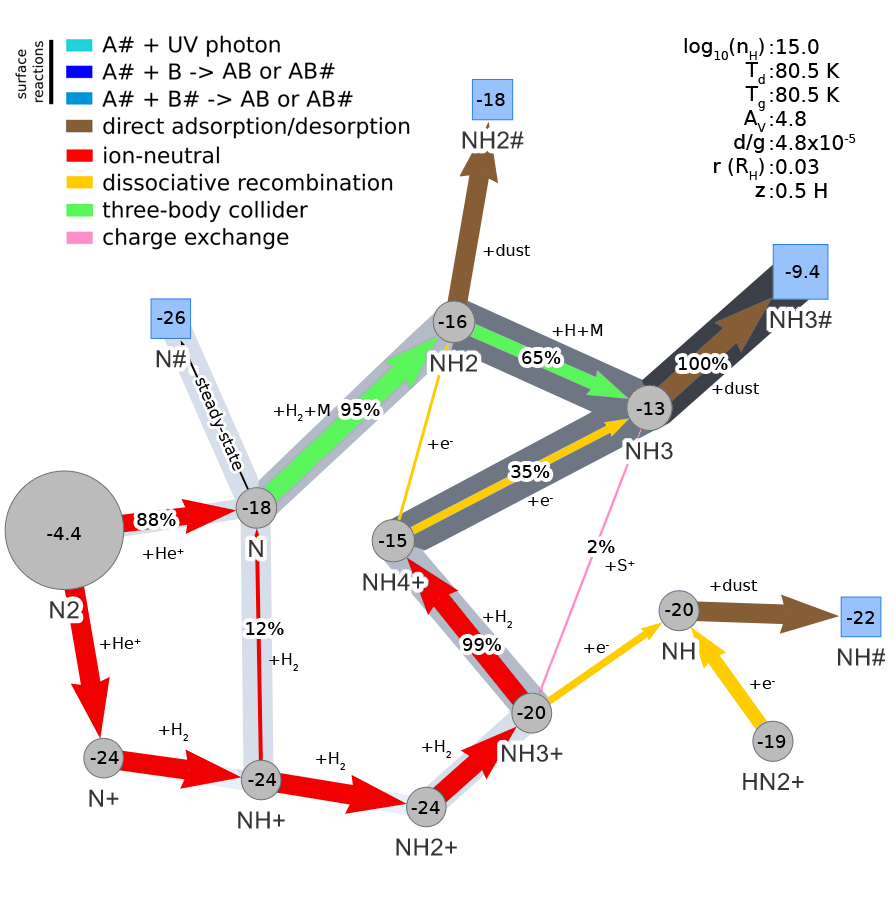}
      \caption{Chemical network diagram illustrating the formation of NH$_3$ ice in the CPD after a partial reset, immediately prior to the accreted gas reaching the midplane. The pathway from N$_2$ to NH3$\#$ is highlighted.  Percentages for reaction A$\rightarrow$B indicate the fraction of reactions forming B which involve species A. A label ``steady-state" indicates that the net rate is zero.}
      \label{fig:net-nh3}
\end{figure}

\noindent 
Here, M = H, H$_2$, or He. The importance of this pathway is illustrated clearly by the green arrows in the chemical network diagram Fig. \ref{fig:net-nh3}. These collider reactions are very efficient at the typical CPD midplane densities (n$_{\rm H} \sim 10^{12}$ cm$^{-3}$).  However, the absence of abundant atomic nitrogen prevents the collider pathway from producing significant quantities of NH$_3$. Then, N$_2$ is destroyed predominantly by reactions with He ions at a relatively low rate, as He$+$ is produced only be cosmic-ray ionization.  The collider pathway to form NH$_3$ thus does not result in significant accumulation of NH$_3$ ice. By the time the gas parcel has reached the midplane, NH$_3$ ice is present only as a trace species.

The collider-pathway begins with the formation of NH$_2$ (Eq. \ref{eq:nh3-collider-path}) which is also relevant to the water formation pathway involving NH$_2$ + O $\rightarrow$ NH + OH \citep{Kamp2017}.  While the pre-exponential factor 10$^{-26}$ cm$^6$ s$^{-1}$ is derived from the work of \citet{Avramenko1966},  we have chosen to adopt a significantly lower rate more typical of collider reactions (10$^{-30}$ cm$^6$ s$^{-1}$), which still produces enough NH$_2$ for this path to be the dominant NH$_3$ formation route in the inner disk.  It has been noted that this particular reaction is critical to accurately reproduce observed OH and H$_2$O gas-phase abundances, but that modern reevaluation of its rate and temperature dependence are needed \citep{Kamp2017}.  For the second collider reaction in this path (Eq. \ref{eq:NH2_collider_2}), we adopted the rate coefficients of \citet{Gordon1971},  listing a pre-exponential factor 6.07 $\times$ 10$^{-30}$ cm$^6$ s$^{-1}$. Other more recent experimental results assuming the reaction to be in the three body pressure regime give values in the range 2.3$\times10^{-30}$ - 1.42$\times 10^{-29}$ for various third bodies \citep{Altinay2012,Altinay2015}, hence, we consider this a reasonable value.

In the outer disk, NH$_3$ gas is efficiently photodissociated.  The NH$_3$ ice is instead formed primarily by barrier-less successive hydrogenation of atomic nitrogen on icy grain surfaces \citep{Charnley2001,Fedoseev2015} which has been experimentally demonstrated to occur \citep{Hiraoka1995,Hidaka2011} via the Langmuir-Hinshelwood mechanism.  The formation pathway is then

\begin{align}   \label{eq:NH2-1}
   &  \rm NH\# + H\# \rightarrow NH_2\#   ,\\ 
   &  \rm NH_2\# + H\# \rightarrow NH_3\#.  \;   \label{eq:NH2-2}
\end{align}

\noindent 
Both in the inner and outer disk NH$_3$ ice does not consitute more than $10^{-3}$ of the total ice by molar fraction.

\subsubsection{Carbon dioxide (CO$_2$)}

\begin{figure}
\centering
  \includegraphics[width=\textwidth/2]{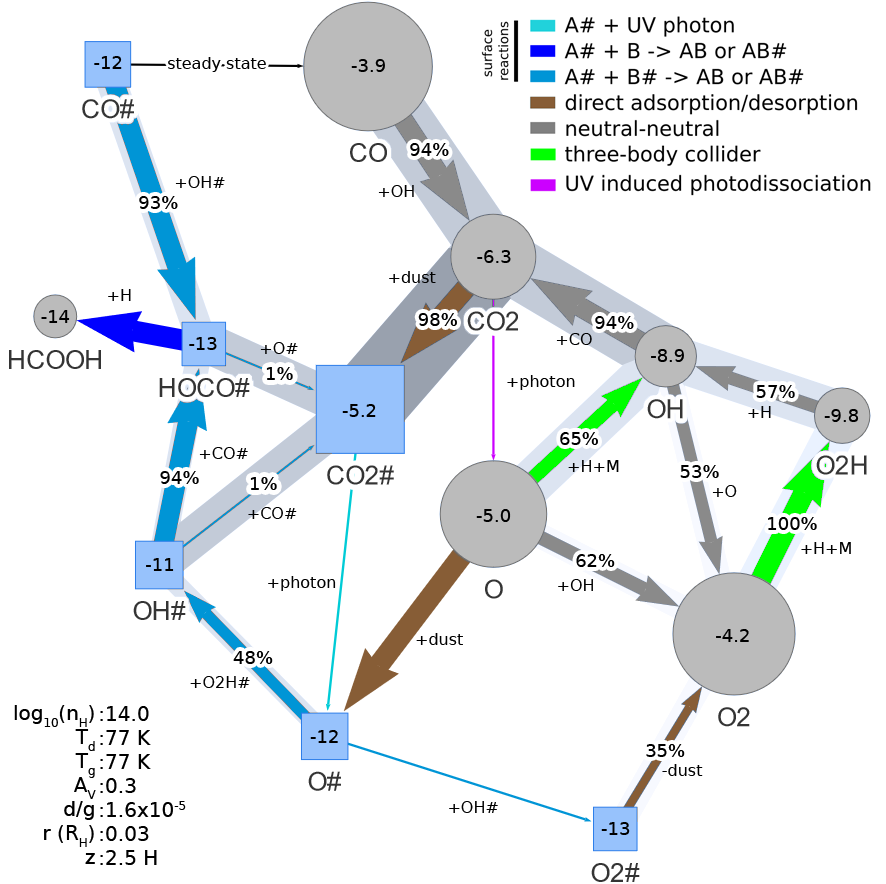}
      \caption{Chemical reaction network illustrating the formation of CO$_2$ ice after a partial reset in which ices accreting onto the CPD are initially sublimated and placed into the gas-phase.}
      \label{fig:net-co2}
\end{figure}

While CO$_2$ ice is initially only a trace species in the accreted circumstellar disk material, it becomes abundant in the CPD  prior to the accreted material reaching the midplane.  The chemical network diagram of the predominant CO$_2$ ice formation paths during this stage can be found in Fig. \ref{fig:net-co2}. This figure illustrates how the production of OH by collider reactions (green arrows) is critical to the efficient formation of CO$_2$ ice.   In the time that accreted gas and ice reside in the optically thin surface layers of the CPD, it initially liberates significant quantities of atomic oxygen from gas-phase H$_2$O, which is hydrogenated via three-body collider reactions.  The OH then reacts with abundant gas-phase CO to produce $98\%$ of the CO$_2$,  which then freezes out onto grains. In particular three-body collider reactions account for nearly all ($>99\%$) OH formation which is critical for the CO+OH gas-phase reaction.  It can also be seen in Fig. \ref{fig:net-co2} that the grain-surface formation of CO$_2$ ice plays only a minor role prior to the gas parcel reaching the midplane.

After the gas and dust parcel reaches the midplane, the chemistry is evolved for an additional 10$^3$-10$^4$ yr for the high- and low-viscosity cases, respectively.  The resulting composition at $t_{\rm visc}$ is similar to that of the full reset case, with the exception that the inner CPD (near the present day orbit of Callisto) also retains a significant CO$_2$ ice component. This can be seen in Fig. \ref{fig:radial-abundances-all} (a) and Fig. \ref{fig:radial-abundances-all} (d). CO$_2$ ice formation continues in the outer CPD at the midplane in the absence of abundant atomic O, as OH is produced instead on grain-surfaces by the photodissociation of H2O$\#$. This is described in the following section and can be seen in Fig. \ref{fig:net-hcooh}. 

\subsubsection{Formic acid (HCOOH)}  \label{sec:results_CO2}

HOCO (hydrocarboxyl radical) and HCOOH (formic acid) are of relevance in the cold, high-density midplane where CO$_2$ ice can form; thus, these were included in our extended chemical network. Formic acid is the simplest carboxylic acid and has been identified in star-forming regions \citep{Schutte1999,Ikeda2001} both in gaseous and solid states, as well as in protoplanetary disks \citep{Favre2018} and in comets \citep{Crovisier2004}.  Its abundance typically varies with 1-10$\%$ of water ice \citep{Bisschop2007}.  

\begin{figure}
\centering
  \includegraphics[width=\textwidth/2]{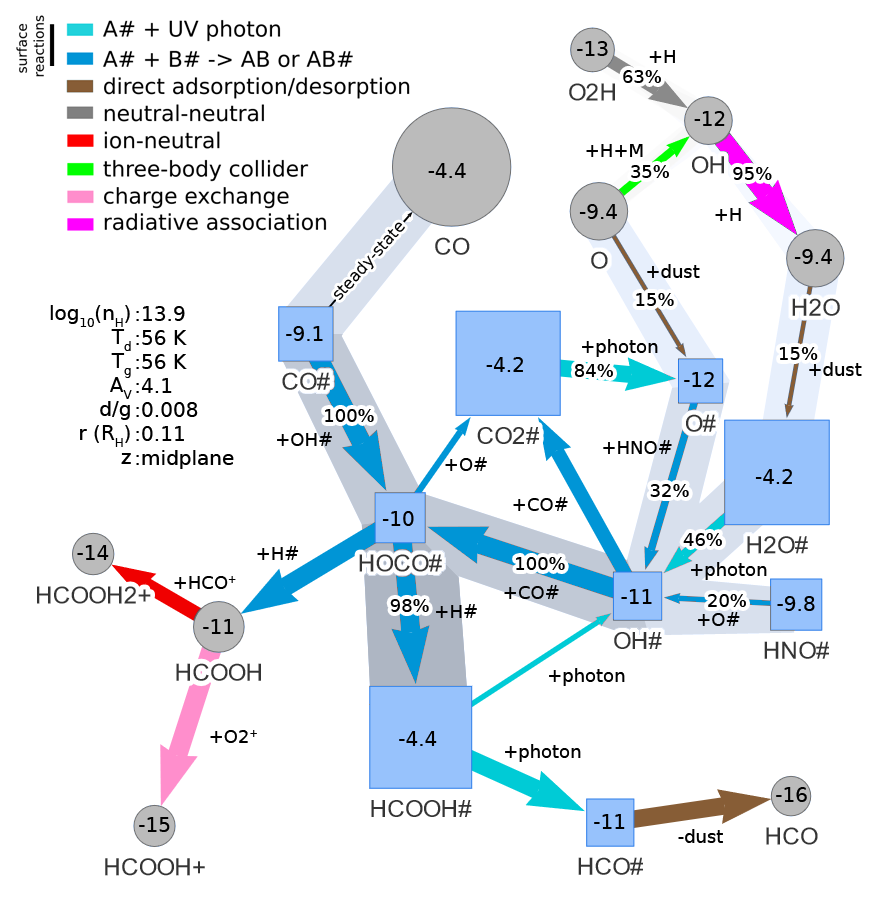}
      \caption{Chemical network diagram centered on the formation of HCOOH ice in the outer regions of the CPD at the midplane.}
      \label{fig:net-hcooh}
\end{figure}

The chemical network diagram of HCOOH formation in the outer CPD can be found in Fig. \ref{fig:net-hcooh}.  It is clear that grain surface reactions play a completely dominant role in this process. In the outer CPD, we find that although it is not stable as an ice, the gas-phase CO freezes out and temporarily occupies a physisorption site on the grain surface. Prior to desorbing the CO$\#$ reacts on the grain surface OH$\#$ to form CO$_2\#$ and H$\#$ \citep{Oba2010,Liu2015}, for which we have adopted the effective barrier of 150\,K \citep{Fulle1996,Ruaud2016}.  

\begin{align}
   & \rm CO + dust   \rightarrow CO\#,        \\
   & \rm CO\# + OH\# \rightarrow CO_2\# + H\#. 
\end{align}
\noindent

Alternatively, as an intermediate step of the OH$\#$ + CO$\#$ reaction the van der Waals complex HOCO$\#$ is formed, which can be hydrogenated to form HCOOH$\#$.

\begin{align}
   & \rm CO\# + OH\# \rightarrow HOCO\#      ,\\  \label{eq:hoco-form}
   & \rm HOCO\# + H\# \rightarrow HCOOH\#. 
\end{align}
\noindent

The HOCO$\#$ formation route can explain the presence of HCOOH$\#$ in cold, dense clouds \citep{Ioppolo2011,Qasim2019}. The resulting radial abundance of HCOOH$\#$ in the reference CPD can be seen in Fig.\ref{fig:radial-abundances-all} (c). In the partial reset case, HCOOH ice can locally constitute a significant fraction of the ices in the reference CPD ($\sim$10mol$\%$). 

We found significant abundances ($\sim10\%$ relative to H$_2$O ice) of HCOOH ice in the outer region of the CPD.  This is comparable to the upper end of inferred abundances ($\sim1-10\%$ level relative to H$_2$O ice) observed toward young stellar objects \citep{Schutte1999,Keane2001,Knez2005,Boogert2015}.  The relatively large abundance of HCOOH ice in the outer CPD relative to its observationally derived abundance in astrophysical ice mixtures in the ISM is noteworthy.  However, this was not entirely unexpected. The minimum CPD temperature set by equilibrium with the background radiation field ensures that a large region in the outer CPD exhibits a narrow range of temperature from 50-55 K.  Given that the majority of the disk surface area is in this zone, the total disk composition is weighted heavily towards these specific conditions.  However, background temperatures as low as 30 K or as high as 70 K do not produce abundant alternative impurities, while the outer CPD remains dominated by CO$_2$ and HCOOH ice.

Additionally, the stability of the HCOOH ice in our model is subject to several uncertainties.  The only grain-surface reaction in our network that is able to destroy HCOOH$\#$ is the photo-induced dissociation to HCO$\#$ and OH$\#$.  Alternatively, it can be placed directly back into the gas phase by thermal, cosmic-ray, or UV-photon induced desorption. We did not include grain-surface hydrogenation of the HCOOH ice. \citet{Bisschop2007} found that hydrogen bombardment of a pure multilayer HCOOH ice does not result in the production of detectable reaction products, concluding that the hydrogenation of HCOOH does not play a role in formation of more complex species and that only minor amounts desorb.  In contrast \citet{Chaabouni2020} found that H-bombardment of a $<1-3$ monolayer coating of HCOOH ice at 10-100 K results in efficient production of CO$_2$ and H$_2$O molecules, as well as CH$_3$OH and H$_2$CO. The authors suggest that this disagreement stems from the inefficiency of H atom diffusion through the pure HCOOH multilayer used in the experimental setup of \citet{Bisschop2007}. Alternatively, the sub-monolayer conditions present in the setup of \citet{Chaabouni2020} potentially cause the substrate itself to become hydrogenated,  increasing the sticking coefficient for H atoms and promoting surface reactions.  Where HCOOH ice is found in our CPD, it has been co-deposited with H$_2$O ice and CO$_2$ ice (with molar ratio H$_2$O:CO$_2$:HCOOH 100:80:80), with an equivalent thickness of several hundred monolayers. Hence we consider it plausible that the majority of the HCOOH embedded within the ice matrix would not be efficiently hydrogenated.

Overall, HCOOH ice has not been detected on the surface of any Galilean moon.  Experimental results indicate that HCOOH ice has a relatively short $8\times10^7$ yr half-life against irradiation by galactic cosmic rays, being dissociated into CO or  CO$_2$ \citep{Bergantini2013}.  Any HCOOH accreted onto the surface of, for instance, Callisto would therefore likely be absent in the present era, having reduced to $<1\%$ of its initial concentration within only 0.56 Gyr.   There is a paucity of research investigating the role of HCOOH in subsurface melts, however, we know that under hydrothermal conditions, water can act as a homogeneous catalyst for the decarboxylation pathway of HCOOH decomposition in liquids \citep{Ruelle1986}, where it decomposes to become the astrobiologically relevant CO$_2$ and H$_2$ molecules \citep{Yu1998}.

\subsection{Full reset (initially atomic gas)}

In the full reset case, the gas in the CPD is initially fully atomic and ionized and no ices are present.  This state represents, for instance, the accretion of a high-mass planet ($M>1M_{\rm J}$), with correspondingly higher infall shock-velocity at the CPD surface, or an accretion of material originating from a greater scale height in the circumstellar disk than we have considered. In the fully reset case, the abundant free atomic hydrogen enables highly efficient combustion chemistry to produce a water-dominated ice composition, as found in \citetalias{Oberg2022}. This efficient water formation locks away the majority of atomic oxygen early on and it is $10^5$ times less abundant than in the partial reset case after 5 yr.  Accordingly, the OH formation rate via O+H is lower and so significantly less OH is available to form CO$_2$ via CO+OH, while the CO abundances are very similar between the two cases (10$^{-3.86}$ vs. $10^{-3.88}$ relative to H$_2$).

Again, ammonia ice is not able to form in abundance as the initially atomic nitrogen is predominantly locked in N$_2$ within a single year via N + NO $\rightarrow $N$_2$ + O or N + NH $\rightarrow $N$_2$ + H. The radial composition of the ices after $10^4$ yr is similar to the partial reset case, although CO$_2$ ice is found in abundance only in the outer disk beyond $\sim2 \times$ the semi-major axis of Callisto. 

In contrast to the partial reset case, the inner disk region is dominated by water ice with a minor ($<1\%$) methanol (CH$_3$OH) component. Methanol is an important primordial solar system volatile and may act as an anti-freeze in subsurface oceans \citep{Deschamps2010,Dougherty2018}.  It has been found to be abundant in solid form near protostars \citep{Dartois1999, Boogert2015}, in comets \citep{Bockelee-Morvan1991,Mumma1993,Bocklee2017,Biver2019}, and in the gas-phase in planet-forming disks \citep{Walsh2016,Booth2021}, where it may be formed via grain-surface reactions involving hydrogenation of CO$\#$ \citep{HIRAOKA1994,Watanabe2002}.  At typical pressures in our reference CPD the freeze-out temperature of methanol is greater than that of NH$_3$ and CO$_2$ \citep{Mousis2009,Johnson2012}. Thus, if the CO$_2$ ice observed on Callisto's surface was formed primordially in the CPD, we could expect that temperatures in the CPD could have allowed for stable methanol ice to be present as well.  Indeed, we find that in the inner disk this occurs for $t_{\rm visc} > 10^3$ yr, where methanol ice is present at the 1$\%$ level at temperatures above 65 K with a peak abundance at 95-100 K.  At these densities, it originates almost exclusively from reactions in the gas-phase via sequential hydrogenation of CO in two- and three-body reactions. Approximately 70$\%$ is formed via:

\begin{align}
   &  \rm CO + H + M  \rightarrow  HCO + M     ,\\
   &  \rm HCO + H + M \rightarrow  H_2CO + M   ,\\ \label{eq:H2CO_hydrogenation}
   &  \rm H_2CO + H   \rightarrow  CH_3O       ,\\ 
   &  \rm CH_3O + H   \rightarrow  CH_3OH,      
\end{align}

\noindent
and the remainder by 
\begin{align} \label{eq:CH2OH-path-1}
    & \rm H_2CO + H    \rightarrow  CH_2OH  ,\\ \label{eq:CH2OH-path-2}
    & \rm CH_2OH + H   \rightarrow  CH_3OH   .
\end{align}

\begin{figure}
\centering
  \includegraphics[width=\textwidth/2]{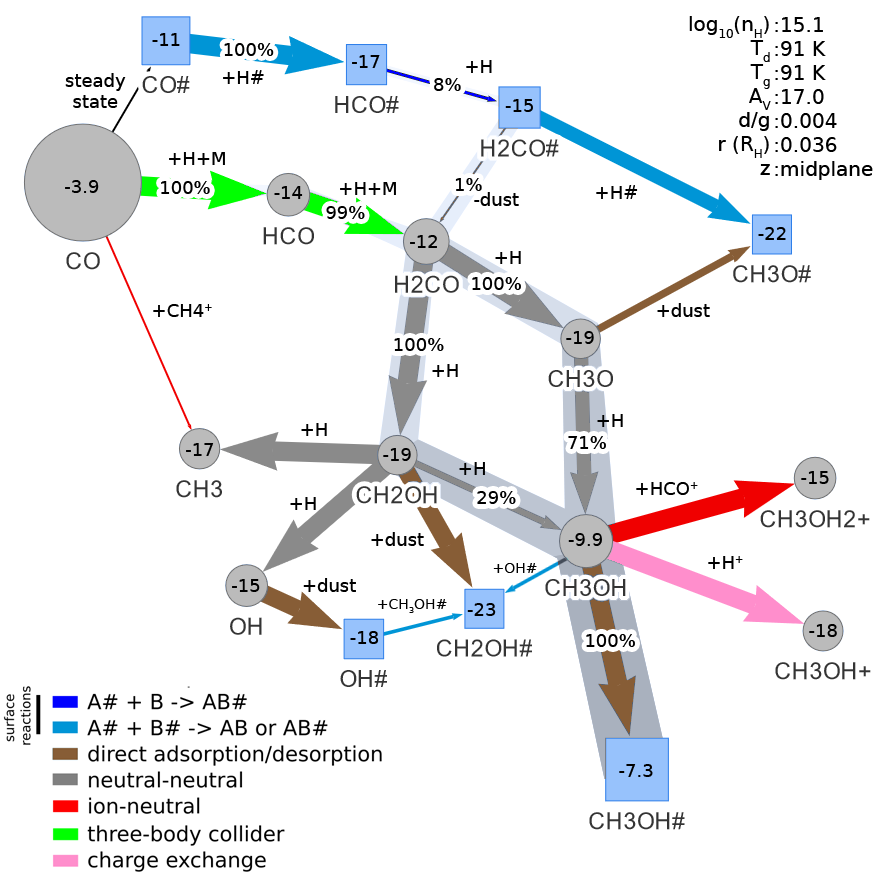}
      \caption{Chemical network diagram centered on the formation of CH$_3$OH at the CPD midplane in the reset case.}
      \label{fig:net-ch3oh}
\end{figure}

\noindent
For the reaction H$_2$CO$\rightarrow$CH$_3$OH, we have adopted the rate coefficients from \citet{Huynh2008} with a barrier of 865 K.  In the absence of this reaction, we find that methanol is produced in similar quantities via the CH$_2$OH pathway.  The rate of formation is thus highly contingent on the availability of free atomic hydrogen in the gas-phase.  The absence of abundant atomic hydrogen prevents the accumulation of methanol in the partial reset or inheritance cases.   An additional "bottleneck" in the reaction network is H$_2$CO. This can be seen in Fig. \ref{fig:net-ch3oh}.  H$_2$CO is formed almost exclusively ($>99\%$) via gas-phase three-body collider reactions.

In the ISM, methanol ice abundances can significantly exceed that which we find in the CPD. 
The grain-surface hydrogenation of H$_2$CO to form CH$_3$O (Eq. \ref{eq:H2CO_hydrogenation}) has been observed at low temperatures experimentally \citep{Hidaka2004,Fuchs2009,Chuang2016}, suggesting that successive hydrogenations of CO can explain the observed abundance of interstellar methanol at low temperatures (< 15 K).  Above this temperature the desorption of H atoms and lower sticking efficiency of H due to the absence of H$_2$ causes a considerable drop in this reaction rate. While these reactions are included in our chemical network, the gas temperature in the CPD does not fall below 50 K; thus, we find  this path to be inefficient.

\subsection{Full inheritance case}  \label{sec:results:full_inherit} 

In the event of a full chemical inheritance from the circumstellar disk gap edge, the ice accreting onto the CPD consists predominantly of water, with ratios H$_2$O:NH$_3$:HCN of 100:15:10 and a significant $\sim10\%$ component of hydrocarbon ices (e.g. C$_2$H$_2$, C$_3$H$_2$). This result is generally consistent with modeling of the outer regions of the circumstellar disk where \mbox{NH$_3$/H$_2$O = 0.14} with as much as 80$\%$ of the nitrogen locked into NH$_3$ and to a lesser extent HCN \citep{Dodson2009}. 

The final composition of the ices in the inheritance scenario is highly contingent on their initial composition. Given the difficulties in correctly capturing the relevant physical conditions at the outer gap edge and the uncertainty from which altitude the gas originates, we consider it more informative to discuss how the ices are altered post-accretion, rather than focusing on their final composition.

Some minor processing of the ices occurs once they are incorporated into the CPD. The more volatile HCN and hydrocarbon ices are lost in the inner region of the disk where only NH$_3\#$ and a minor component of hydrocarbon ices remain as impurities. In the outer region of the CPD, some conversion of H$_2$O and HCN to HCOOH occurs, and to a minor extent CO$_2$.  

At temperature below 70 K HCOOH is co-deposited with the NH$_3$ ice.  In the presence of the proton acceptor NH$_3$, HCOOH will convert to the formate anion HCOO$^-$ and NH4$^+$   \citep{Hudson1999,Schutte1999,Galvez2010} however formate is not included in our chemical network. Likewise the salt formation reaction is not included in the network. We consider what impact the inclusion of this process could have on our final derived abundances.  While the activation barrier of the reaction is negligible, the barrier against diffusion prevents it from occurring at 50-70 K \citep{Schutte1999}.  However some of the HCOOH will react immediately upon deposition due to the acid and base being in direct contact at adjacent binding sites. 10$\%$ of the HCOOH ice is observed to react promptly at 10 K in H$_2$O-NH$_3$-HCOOH mixtures with equal parts NH$_3$-HCOOH \citep{Schutte1999}. Hence we might expect that as much as $\sim20\%$ of the HCOOH present in the outer disk could be converted upon adsorption to HCOO-NH$_4$+.

\subsection{Differing diffusion timescales}

Owing to the uncertainity in the diffusion timescale on which the gas parcel drifts towards the CPD midplane we considered also the case of 10$\times$ shorter and longer $t_{\rm diff}$.  For $t_{\rm diff} = 100$ yr all three initial conditions converge towards a similar final composition which is  CO$_2$-dominated ($>95\%$ by weight) across the entire disk. This is clearly inconsistent with observations of the Galilean moons.  The shorter $t_{\rm diff} = 1$ yr leaves the chemistry less affected by the time spent at the disk surface.  In the partial reset case, a minor fraction (3$\%$) of the accreted circumstellar NH$_3$ ice survives and can still be found at the CPD midplane after $10^4$ yr. In the full reset case, the CH$_3$OH component in the inner disk region becomes more substantial, increasing to a peak of 4$\%$ of the total ice mass.  This additional CH$_3$OH forms because more of the initially atomic hydrogen survives until ices become stable against photodissociation, and are available to hydrogenate H$_2$CO and CH$_3$O.

\section{Implications}

\subsubsection*{Absence of ammonia as an indicator of chemical reset}

We have found that a partial or complete chemical reset of the CPD tends to suppress NH$_3$ formation as efficient N$_2$ self-shielding locks up nitrogen in N$_2$. Even if a substantial component ($\sim20-30\%)$ of NH$_3$ ice were present in Jupiter's feeding zone, a partial or complete reset would prevent its accumulation in the building blocks of the moons. Without a substantial NH$_3$ component the liquidus temperature of the Galilean subsurface oceans may not differ substantially from that of a pure water ice.   
Europa appears to be the only Galilean moon where tectonic or cryovolcanic processes have recently exchanged material between the surface and subsurface where it could provide clues to composition of an ocean  \citep{Kargel2000,Zolotov2001}.   NH$_3$ brought to the surface in the form of an NH$_3$-H$_2$O matrix could be lost on geologically brief timescales to external radiation \citep{MOORE2007260,Bergantini2014}. Longevity of surface ammonia might be extended if it would appear in a more stable form such as a hydrate or salt \citep{COOK201830} but no positive detection has thus far been made \citep{Clark2014}. The non-detection of ammonium compounds on Europa's surface is compatible with a lack of ammonia in a subsurface ocean, although is certainly not conclusive evidence of its absence.

In contrast to the Galilean system, several lines of evidence indicate the presence of NH$_3$ ice during the accretion of the Saturnian moons. The inferred interior composition of the Saturnian moon Enceladus appears to resemble more closely well-mixed outer solar system material and is generally consistent with a composition inherited from solar nebular (cometary) material \citep{Waite2009}.  Enceladus contains a liquid water ocean \citep{Thomas2016} from which interior material is ejected through plumes \citep{Spahn2006, Porco2006, Waite2006}.  The presence of $\mathrm{NH_{3}}$ in the plumes of Enceladus has been established by measurements from several instruments onboard the Cassini spacecraft \citep{Waite2009}  at $> 0.1 \%$ relative to $\mathrm{H_{2}O}$, besides $\mathrm{CO_{2}, CH_{4} \: and \: H_{2}}$  \citep{magee2017neutral}. Likewise NH$_3$ ice is considered to be a likely source of Titan's nitrogen \citep{McKay1988,Sekine2011,Mandt2014}.

We suggest that the CPDs of sufficiently massive planets lose accreted NH$_3$ ice to mild accretion shocks and subsequent chemical evolution, and that the absence of NH$_3$ ice may indicate a (partial) chemical reset has occurred.  As NH$_3$ represents one of the most potent and potentially abundant anti-freezes, subsurface ocean occurrence rates and longevity may then be relatively enhanced in the icy moons that accompany lower-mass giant planets which inherit circumstellar material.

\subsubsection*{Carbon Dioxide at the origin of Ganymede and Callisto}


Several lines of evidence suggest the surface of Callisto is among the most primordial of the regular satellites, potentially providing a direct link to the formation environment of the Galilean moons \citep{Moore2004,Zahnle2003}.  CO$_2$ ice has been detected on the surface of both Ganymede and Callisto \citep{Carlson1996,McCord1997} but only appears below the surface on Callisto \citep{Hibbitts2002,Hibbitts2003}, where it appears to be exhumed by impact cratering. In contrast, CO$_2$ on the surface of Ganymede appears to be of exogeneous or radiolitic origin \citep{Hibbitts2003}. Hence if we consider Callisto's reservoir of CO$_2$ ice to be primordial we can consider which of our assumptions are consistent with its presence.

In the partial reset case, which we considered to be a priori the most likely initial condition  of accreted material,  CO$_2$ ice is present in significant quantities at the present-day position of Callisto but less so near Ganymede. Superficially this appears to be consistent with the proposed distinct origins of Ganymede and Callisto's CO$_2$.  However, the local ice mass fraction of CO$_2$ in the CPD is high ($\geq$60$\%$). This appears to be in conflict with the inferred surface abundance of CO$_2$ ice on Callisto, where it constitutes no more than 0.01-0.16$\%$ of the host material mass \citep{Hibbitts2002}.  It is however unclear whether the observationally inferred surface abundance of CO$_2$ on Callisto is truly representative of the subsurface composition.  Pure CO$_2$ ice is not stable at the surface of the Galilean moons and CO$_2$ may instead be present in the form of clathrates \citep{CHABAN2007}. Hence, an initially large CO$_2$ component exposed to the surface could have been lost to sublimation and dissociation. A substantial subsurface CO$_2$ reservoir is nevertheless implied, given the continuous replenishment of Callisto's CO$_2$ exosphere \citep{Carlson1999b}.  In contrast to the partially reset case,  we find CO$_2$ ice at a concentration of $\sim0.2\%$ near Callisto's location in the fully reset CPD. While this appears to be more representative of what is known of the Galilean moon surface composition, the primordial CO$_2$ concentration of Callisto's building blocks cannot simply be derived from the present state of the surface.

Our findings are consistent with a primordial origin for Callisto's CO$_2$, and point to the possibility that Ganymede and Callisto's icy building blocks had distinct chemical compositions.  While it has been suggested that Ganymede may have formed with a primordial CO$_2$ component which was lost during an episodic period of surface melting, our results suggest icy grains in its vicinity were CO$_2$-poor.  A CPD midplane temperature profile which is dominated by viscous heating and in which the water iceline falls between Europa and Ganymede naturally produces a CO$_2$ iceline between Ganymede and Callisto.

\section{Summary and Conclusions}

If CPD ice composition is (partially or fully) reset, NH$_3$ ice formation is inefficient due to N$_2$ self-shielding.  The resulting $\ll$1$\%$  concentration of NH$_3$ ice is unlikely to significantly alter the thermophysical/chemical properties of subsurface melt.   The most significant impurities are the carbon-bearing  CO$_2$ and HCOOH ices and each make up at most $\sim10\%$ of the molar ice fraction.  If the growth of the Galilean moons occurred near their present-day positions they are largely free of impurities, being composed of 98$\%$ water ice, $\sim2\%$ CH$_3$OH, and trace amounts of CO$_2$.   If instead the CPD ice composition is inherited from the circumstellar nebula, NH$_3$ ice can survive conditions at the CPD midplane and becomes the most abundant impurity. Observations indicating the presence of NH$_3$ in the Saturnian satellite system but not in the Galilean one are consistent with a reset-inheritance dichotomy. NH$_3$ in the planetary feeding zone of Jupiter, if present, may have been destroyed during accretion onto the CPD and then could not form again in time.  Our key findings are summarized as follows: 

\begin{enumerate}

\item The ice composition of the Galilean moons corresponds to a partial or full chemical reset, as opposed to the ices of the Saturnian moons, which may have been more directly inherited from the circumstellar disk. 

\item  A partial reset prevents efficient formation of ammonia ice. The building blocks of the Galilean moons (and of exomoons forming in similar CPDs) would be  nitrogen-poor (NH$_3$ ice abundances with respect to the H$_2$O ice of $\sim0.1\%$). 

\item Our results are consistent with a primordial origin for CO$_2$ ice on Callisto and an ice composition that is chemically distinct from Ganymede.

\end{enumerate}

The composition of the building blocks that form moons around giant planets is determined by the conditions of accretion onto the planet's CPD, which in turn is influenced by the mass and orbital properties of the planet. The compositional reset-inheritance dichotomy of CPD ices ties together the properties of the planet and the long-term geophysical evolution and composition of icy satellite interior oceans.

\begin{acknowledgements}
The research of N.O. and I.K. is supported by grants from the Netherlands Organization for Scientific Research (NWO, grant number 614.001.552) and the Netherlands Research School for Astronomy (NOVA). This research has made use of NASA's Astrophysics Data System Bibliographic Services. This research has also extensively used Numpy \citep{numpy},  Matplotlib \citep{matplotlib}, Scipy \citep{scipy}, and Prodimopy \url{https://gitlab.astro.rug.nl/prodimo/prodimopy}. N.O would like to thank S. Ceulemans for her suggestion that greatly improved the visualizations in this work, as well as J. Tjoa and S. van Mierlo for helpful discussions and support. 
\end{acknowledgements}


%

\bibliographystyle{aa} 
\bibliography{refs.bib} 

%

%


\begin{appendix} 

\section{Chemical network diagrams annotated} \label{appendix:diagrams}

The chemical network diagrams are generated algorithmically according to the following rules:

\begin{enumerate}
    \item A single species is selected as the seed from which the diagram will be generated. 
    \item In the first iteration, all species which are involved in the formation or destruction of the seed above a certain rate threshold are added as nodes to the network (one degree of separation).  
    \item The reaction with the highest rate relating two species is represented as a connection (arrow) between the two nodes. 
    \item This process is iterated starting from (2) for $n$ degrees of separation.
\end{enumerate}

\begin{figure}[ht]
  \includegraphics[width=\textwidth/2]{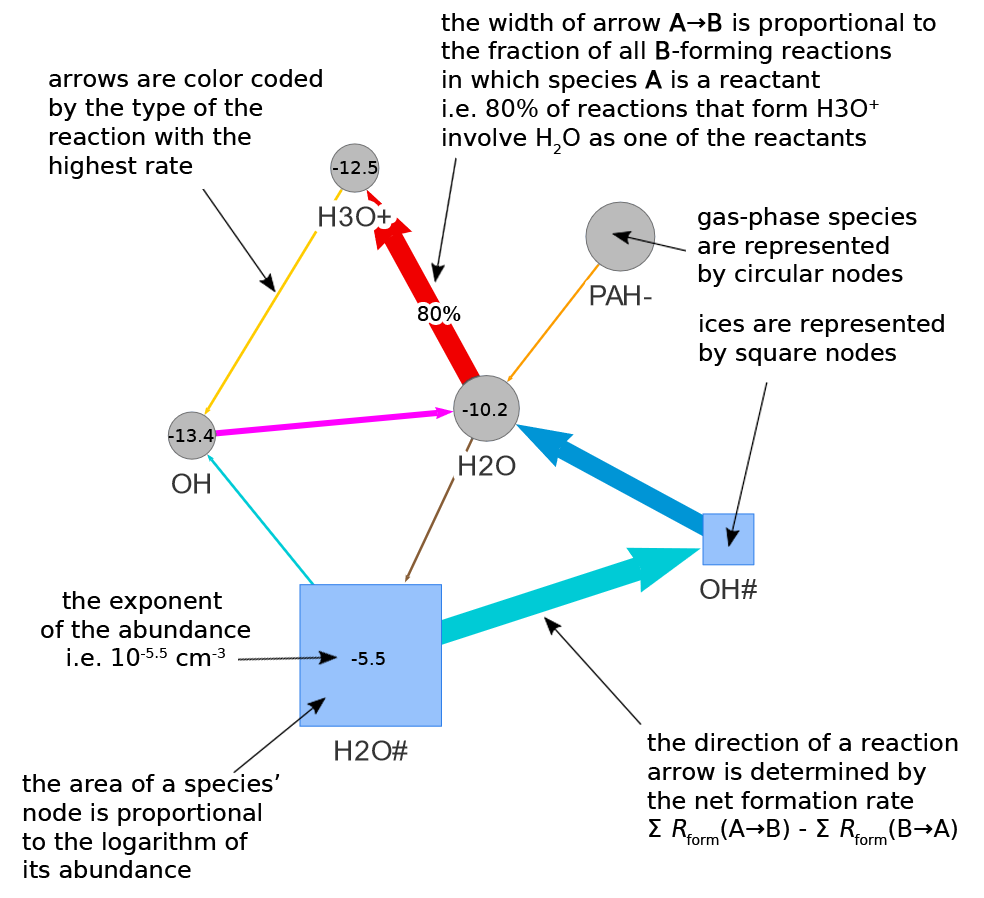}
      \caption{Example chemical network diagram with annotations describing how the diagram can be interpreted. The primary formation pathway of a species may also be highlighted for reference (not depicted).}
      \label{fig:diagram-primer}
\end{figure}

An annotated example chemical network diagram can be found in Fig. \ref{fig:diagram-primer}.  While one species may be involved in the formation of another through multiple distinct reactions (i.e. H+OH$\rightarrow$H$_2$O and  OH+OH$\rightarrow$H$_2$O+O), we color code the reaction connecting the nodes with the type (i.e. ion-neutral) of the reaction with the highest rate $v$.  The arrow width is proportional to the net formation rate ($v$(A$\rightarrow$B) - $v$(B$\rightarrow$A)) between the two species.  The node size is proportional to the abundance of the species.  For some reactions the fraction of the reactions forming species $B$ involving species $A$ is included as a percentage (i.e. for A$\rightarrow$B 20$\%$ of all the reactions that form B involves A).  Networks are generated for a single grid point in the CPD model (ambient conditions are listed in the respective diagram) and at a specific time only, and are thus to be considered as ``snapshots".   H and H$_2$ are not displayed as nodes. Arrows are not displayed if they contribute less than 1$\%$ to the total formation rate of a species.  

\section{Initial chemical abundances} \label{appendix:initial}

The top 10 most abundant species for each chemical initial condition can be found in Table \ref{tab:abunds} normalized to the most common species.

\begin{table}[ht] 
\centering

\caption{Abundance* of the 10 most common species for each initial condition.} 

\begin{tabular}{lr|lr|lr}  
\multicolumn{2}{c|}{\bfseries Reset} & \multicolumn{2}{|c|}{\bfseries Partial Reset} & \multicolumn{2}{|c}{\bfseries Inherit} \\  \hline 

H+    & 0.00   & H2      &  0.00 &   H2      & 0.00  \\
He+   & -1.02  & He      & -0.71 &   He      & -0.71 \\
O++   & -3.52  & H2O     & -3.43 &   H2O\#   & -3.43 \\
C++   & -3.86  & CO      & -3.63 &   CO      & -3.63 \\
N++   & -4.10  & Ne      & -3.74 &   Ne      & -3.74 \\
S++   & -6.73  & NH3     & -3.81 &   NH3\#   & -3.81 \\
Si++  & -7.76  & C2H4    & -4.68 &   C2H4\#  & -4.68 \\
Mg++  & -7.97  & NH2     & -5.32 &   NH2\#   & -5.32 \\
Na++  & -8.64  & O       & -5.61 &   O       & -5.61 \\
Fe++  & -8.76  & Ar      & -5.61 &   Ar      & -5.61
    \end{tabular}
    
     {\raggedright * Abundances are expressed as the log$_{10}$ of the abundance relative to the most abundant species. \par}

\label{tab:abunds}
\end{table} 

\vspace{-2ex}
\section{Freeze-out of elements} \label{appendix:freezeout}

\begin{figure}[ht]
  \includegraphics[width=\textwidth/2]{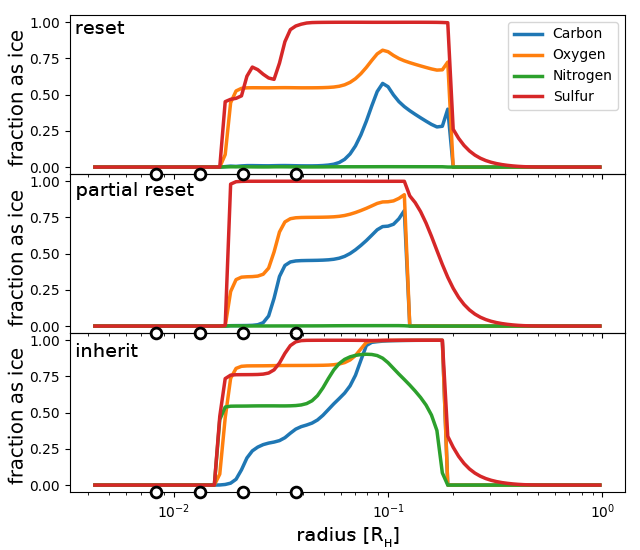}
      \caption{The fraction of nuclei of various elements which are found in the ice at the midplane of the CPD for the three initial conditions.  Properties have been extracted from the low viscosity CPD with the viscous timescale of $10^4$ yr. Radius is in units of the Jupiter Hill radius.  The four empty circles indicate the position of the Galilean satellites.}
      \label{fig:f_element_ice}
\end{figure}

While several of the ices appear at the CPD midplane only as trace species (e.g. SO$_2$) the freeze-out of the corresponding elements may be complete.  The fraction of carbon, oxygen, nitrogen, and sulfur nuclei which are found as ice at each radius in the CPD midplane at $t_{\rm visc}=10^4$ yr for the reset, partial reset, and inheritance cases can be found in Fig. \ref{fig:f_element_ice}.  In all cases Sulfur is found to have frozen-out of the gas almost entirely where possible. This is consistent with the origin of Io's sulfur being in the form of refractory iron sulfides, rather than as e.g. SO$_2\#$, as we find it co-deposited with H$_2$O$\#$ in all cases \citep{Carlson2007}.

\end{appendix}

\end{document}